\documentclass[useAMS,usenatbib]{mn2e}
\usepackage{times}
\usepackage{graphicx} 
\def\ltsima{$\; \buildrel < \over \sim \;$}
\def\lsim{\lower.5ex\hbox{\ltsima}}
\def\gtsima{$\; \buildrel > \over \sim \;$}
\def\gsim{\lower.5ex\hbox{\gtsima}}
\def\ga{\mathrel{\hbox{\rlap{\hbox{\lower4pt\hbox{$\sim$}}}\hbox{$>$}}}}
\def\la{\mathrel{\hbox{\rlap{\hbox{\lower4pt\hbox{$\sim$}}}\hbox{$<$}}}}

\title[Time-evolution of ionization and heating around first stars and
miniquasars]{Time-evolution of ionization and heating around first
  stars and miniqsos}
\author[Rajat M. Thomas and Saleem Zaroubi]{Rajat M. 
Thomas\thanks{thomas@astro.rug.nl} and Saleem Zaroubi\thanks{saleem@astro.rug.nl}\\
  Kapteyn Astronomical Institute, Landleven 12,Groningen 9747 AD, The
  Netherlands}
\begin{document}

\date{}

\pagerange{\pageref{firstpage}--\pageref{lastpage}} \pubyear{2007}

\maketitle

\label{firstpage}

\begin{abstract}
  
A one dimensional radiative transfer code is developed to track the
ionization and heating pattern around the first miniquasars and
Population III stars.  The code follows the evolution of the
ionization of the species of hydrogen and helium and the intergalactic
medium temperature profiles as a function of redshift.  The radiative
transfer calculations show that the ionization signature of the first
miniquasars and stars is very similar yet the heating pattern around
the two is very different. Furthermore, the first massive miniquasars
($ \gsim 10^5 M_\odot$) do produce large ionized bubbles around them,
which can potentially be imaged directly using future radio
telescopes. It is also shown that the ionized bubbles not only stay
ionized for considerable time after the switching off of the source,
but continue to expand for a short while due to secondary collisions
prompted by the X-ray part of their spectra.  Varying spectral shapes
also produced sizable variations in ionized fraction and temperature
profile. We also compare the radiative transfer results with the
analytical approximation usually adopted for heating by miniquasars
and find that, because of the inadequate treatment of the He species, the
analytical approach leads to an underestimation of the temperature in the
outer radii by a factor $\approx 5$.  Population III stars -- with
masses in the range of $10 - 1000~M_\odot$ and modelled as blackbodies
at a temperature of $50000$~K -- are found to be efficient in ionizing
their surroundings.  The lack of very high energy photons limits the
extent of heating of these first stars and have a distinctly different
signature from that of the miniqsos. Observational effects on the 21 cm
brightness temperature, the thermal and kinetic Sunyaev-Ze'ldovich
effects, are also studied in the context of the upcoming radio and
microwave telescopes like LOFAR and SPT.

\end{abstract}


\section{Introduction}

Comprehending the end of the so called ``dark ages of the Universe"
through the process of reionization constitutes a very important step
in our understanding of the formation and build up of structure in the
Universe. It also provides a powerful constraint on various cosmological
models \citep{tozzi,bharadwaj,zal}. Observation of this crucial epoch
in the history of the Universe is possible through the
redshifted 21 cm emission from neutral hydrogen as first recognized by
\citet{sz75} and further improved upon and developed by many authors
like \citet{hogan}, \citet{scott} and \citet{madau}.

One of the popular views on how reionization occurred is the following
\citep{fur_a}; first sources of light, be it Population III
(hereafter, Pop III) stars, miniquasars or any other other source of
radiation, created ionized bubbles. These ionized bubbles probably
started to form around high density peaks, roughly at the same time,
everywhere in the Universe. As the sources number of increased and
since the expansion rate of these bubbles was greater than the
Universe's expansion, the bubbles overlapped around a certain
redshift. A novel class of large radio telescopes telescopes like
LOFAR\footnote{www.lofar.org}\footnote{www.astro.rug.nl/\~{
}LofarEoR}, MWA\footnote{http://www.haystack.mit.edu/ast/arrays/mwa/},
21CMA\footnote{http://21cma.bao.ac.cn/} and
SKA\footnote{www.skatelescope.org} are being designed, among other
things, to map the 21 cm emission from neutral diffuse inter galactic
medium (IGM) at the Epoch of Reionization (EoR) as a function of
redshift, where they will detect the statistical signature of the high
redshift Universe as a function of z \citep{zal,fur_b}. The ultimate
goal of these mega radio telescopes would lie in resolving and mapping
the ionized bubbles and their evolution around the first sources of UV
radiation.

Recent observations show that quasars with black hole masses as large
as $10^9 M_\odot$ have already existed at redshift 6 and higher
\citep{fan03,fan06} suggesting the existence of miniquasars that
harbor intermediate mass black holes with masses in the range of $10^{3-6}
M_\odot$ at very high redshifts ($z>10$). A scenario in which a black
hole starts accreting mass at very high redshift thus producing UV and
X-ray radiation that ionizes and heats up its surrounding IGM has been
studied by many authors
\citep{ricottia04,ricottib04,nusser05,fur_b,wyithe,fur_l,dijkstra}.
Many aspects of the above scenario are debatable, e.g., what was the
primordial black hole formation mechanism \citep{begel,spaans06}? and
what was the evolution rate of their mass densities as a function of
redshift which varies dramatically depending on the model used (c.f.,
\citet{begel}, \citet{rhook} and \citet{zaroubi07} for different
scenarios)? Furthermore, the spectral indices of the power law
spectral energy distribution of a quasar, even if marginally
different, can suggest very different imprints. Also the accretion
rate and lifetimes of these primordial objects are completely unknown.

Apart from these large potential variations in parameters within the
scenario of miniqsos, we also have the distinctively different
scenario of Pop III stars as the primary source of reionization
\citep{benson06,whalen,abel07,ciardi,barkana01,loeb06,miralda,bromm01,
kitayama,wyithe03}.
Within the context of Pop III stars there are many uncertainties that
need to be considered. For example, the unknown blackbody temperature
of the star (assuming they can be characterized by one), its mass,
life-times, clustering properties and redshifts at which they start to
appear.

In this paper, we try to fully explore the implications of varying the
above mentioned parameters on the final maps of the 21 cm signal for
individual objects that are initially surrounded by neutral IGM. In
order to achieve a full sample of this parameter space it becomes
imperative that a fast and realistic method of modelling these
parameters is introduced. Previous simulations of radiative transfer
in cosmological context and studies on the impact of first sources on the 
surrounding IGM, show that the regions that are ionized around
these first sources, are almost spherically symmetric, e.g., Figure 7
of \citet{kuhlen}.  Thus we resorted to developing and using a 1-D
radiative transfer code to study the ionizing fronts (I-front), their
velocity, size, shape and so on. 

Quick execution times and the simplicity in setting up the initial
conditions allow us to simulate the influence of the miniquasars with
varying masses for different redshifts, lifetimes and spectral
indices.  Specifically, we explore the influence of individual black
hole masses ranging from $ 100 M_\odot $ to $ 10^8 M_\odot$, redshifts
between $10$ and $30$ and lifetimes in the range of $3$ to $20$
million years. The influence of first stars on the IGM was explored
with mass ranges of $10 M_\odot$--$10^3 M_\odot$, whose luminosity is
determined mainly by their mass\citep{schaerer}.

The upcoming radio telescopes like LOFAR, MWA, 21CMA and SKA will be
used to observe the redshifted 21 cm from the dark ages and the epoch of
reionization. Hence, trying to predict the topology, intensity and
distribution of the ionized regions as a function of redshift is
crucial in the development of sophisticated data analysis techniques
to retrieve the 21 cm EoR signal. Given the spatial resolution of
these radio telescopes, it is very hard to differentiate between the
ionization bubbles around miniqsos and Pop III stars. \citet{miralda,
kuhlen, zaroubi05} have suggested ways to differentiate between
various sources of ionization.  In this paper it is shown that the
heating caused by these two possible ionization sources is
significantly different especially at the early stages of the EoR, 
leading to different strengths of coupling between the
spin and kinetic temperature which is reflected in the measured
brightness temperature. Therefore we argue that this feature could, in
principle, be used to distinguish between ionization sources.

 The paper is structured as follows: In section \S\ref{simulations} we
detail the rate equations used in the simulations and also describe
how we solve for the radial dependencies of the ionized fractions of
the various species involved and of the temperature.  Accuracy and
reliability of the radiative transfer code developed were
tested. Results of the two tests conducted, namely, propagation of the
ionizing front into the mean non-expanding IGM and the Shapiro \&
Giroux case in which the IGM is expanding with the Universe, are
presented in \S\ref{tests}. \S\ref{miniqso} outlines some of the
results obtained in the application of this code to power-law
sources. A similar exercise is carried out for the case of stars and
results discussed in \S\ref{stars}. The coupling mechanisms between
the spin and kinetic temperatures and its effect on the brightness
temperature is discussed in \S\ref{observational}.  Other observables
like the thermal and kinetic Sunyaev-Ze'ldovich effects caused by
these sources are considered in \S\ref{szsection}. We present our
conclusions and the possibilities of expanding this work further in
\S\ref{conclusions}. We also present a comparison of the heating
profiles provided by the code developed here with an analytical
approach developed earlier in the appendix \S\ref{analytical}.

\section[]{Simulations} \label{simulations}

The results presented in this study are based on a 1-D radiative
transfer (RT) code with which the evolution of HI,
HII, HeI, HeII, HeIII, free electrons and temperature is monitored
in time. In order to achieve this, we follow \citet{fuku}, wherein a
set of coupled differential equations (\ref{xhfrac} to \ref{xhe3frac}
and \ref{eq:temp}) are solved . We follow \citet{jones} and 
\citet{fuku} in treating hydrogen as two-level system plus a continuum.

For this study, in the case of miniqsos, we simply assumed
that the density surrounding it is the mean density in an expanding
Universe. For the case of stars, we assume an isothermal like
profile of the form $\rho(r) ~ [\mathrm{cm}^{-3}] = 3.2 \times (91.5 ~
\mathrm{pc}/r)^2$, as in \cite{mellema}, until the density reaches the mean
IGM density, after which we set the density to that of the mean
IGM. The cosmological parameters used here are those set by the WMAP
3$^{rd}$ year results \citep{spergel07}.

\subsection{Rate equations}

In the RT code we developed, the radiation is allowed to propagate
radially with an underlying density profile, which in our case is the 
constant background IGM density. The code solves the
following rate equations and finds the number density of each
species in time and space:

\begin{eqnarray}\label{xhfrac}
\frac{\mathrm{d n_{H_{II}}}}{\mathrm{dt}} &  = & \mathrm{\Gamma_{H_I} n_{H_I}-\alpha_{H_{II}} n_e n_{H_{II}}}
\end{eqnarray} 

\begin{eqnarray}\label{xhe2frac}
\frac{\mathrm{d n_{He_{II}}}}{\mathrm{dt}} &  =  & \mathrm{\Gamma_{He_{I}} n_{He_{I}}+ \beta_{He_I} n_e n_{He_I}} \nonumber \\ 
& & - \mathrm{\beta_{He_{II}} n_e n_{He_{II}} - \alpha_{He_{II}} n_e n_{He_{II}}} \nonumber \\
& & + \mathrm{\alpha_{He_{III}} n_e n_{He_{III}} - \xi_{He_{II}} n_e n_{He_{II}}} 
\end{eqnarray} 

\begin{eqnarray}\label{xhe3frac}
\frac{\mathrm{d n_{He_{III}}}}{\mathrm{dt}} &  =  & \mathrm{\Gamma_{He_{II}} n_{He_{II}}+ \beta_{He_{II}} n_e n_{He_{II}}} \nonumber \\ 
& & - \mathrm{\alpha_{He_{III}} n_e n_{He_{III}}} 
\end{eqnarray}

\begin{eqnarray}\label{gammah1}
 \lefteqn{\mathrm{\Gamma_{H_I}}  =  \gamma_{2c} + \beta_{H_I} n_e + \int\limits_{\rmn{E_{H_I}}}^\infty \sigma_{H_I} N(E;r;t) \frac{\rmn{d}E}{E}+ }\nonumber \\
& & {} f_H \left [ \int\limits_{\rmn{E_{H_I}}}^\infty \sigma_{H_I} \left ( \frac{E-E_{H_I}}{E_{H_I}} \right ) N(E;r;t) \frac{\rmn{d}E}{E} \right ]  + \nonumber \\
& & {} f_H \left [ \frac{n_{He_{I}}}{n_{H_{I}}} \int\limits_{\rmn{E_{He_I}}}^\infty \sigma_{He_I} \left ( \frac{E-E_{He_I}}{E_{H_I}} \right ) N(E;r;t) \frac{\rmn{d}E}{E} \right ]
\end{eqnarray}

\begin{eqnarray}\label{gammahe1}
 \lefteqn{\mathrm{\Gamma_{He_I}}  =  \int\limits_{\rmn{E_{He_I}}}^\infty \sigma_{He_I} N(E;r;t) \frac{\rmn{d}E}{E} + }\nonumber \\
 & &  {} f_{He} \left [ \int\limits_{\rmn{E_{He_I}}}^\infty \sigma_{He_I} \left ( \frac{E-E_{He_I}}{E_{He_I}} \right ) N(E;r;t) \frac{\rmn{d}E}{E} \right ]  + \nonumber \\
 & &  {} f_{He} \left [ \frac{n_{H_{I}}}{n_{He_{I}}} \int\limits_{\rmn{E_{He_I}}}^\infty \sigma_{H_I} \left ( \frac{E-E_{H_I}}{E_{He_I}} \right ) N(E;r;t) \frac{\rmn{d}E}{E} \right ] 
\end{eqnarray}

\begin{equation}\label{gammahe2}
   \mathrm{\Gamma_{He_{II}}}=\int\limits_{\rmn{E_{He_{II}}}} \sigma_{He_{II}} N(E;r;t) \frac{\rmn{d}E}{E}
\end{equation}
 
where, $\rmn{n_{H_I}, n_{H_{II}}, n_{He_I}, n_{He_{II}}}$ and
$\rmn{n_{He_{III}}}$ are the neutral \& ionized hydrogen, neutral,
single and doubly ionized helium densities respectively. $\rmn{n_e}$
is the total electron density given by $\rmn{n_e} =
\rmn{n_{H_{II}}}+\rmn{n_{He_I}}+2
\rmn{n_{He_{II}}}$. $E_{H_{I}}$, $E_{He_{I}}$ and $E_{He_{II}}$
are the ionization energies for the corresponding species .$\beta$
are the collisional ionization coefficients, $\rmn{\alpha}$ and
$\rmn\xi$, the recombination coefficients, $\rmn\sigma$ are the
bound-free photo-ionization cross-sections,
$\rmn\gamma_{2c}=\alpha_{H_I} (T_\gamma) \times
(m_ekT_\gamma/2\pi)^{3/2}e^{-3.4eV/T_\gamma}$, is the photoionization
coefficient due to background photons. The subscripts denote the
species to which the coefficient belongs.

The radiation flux $N(E;r;t)$, is the same as in \citet{zaroubi05};
\begin{equation} \label{numberdensity}
N(E;r;t) = e^{-\tau (E;r;t)} \frac{Ag}{(r/Mpc)^2}I(E) \quad \left [\rmn{cm}^{-2} \rmn{s}^{-1}\right ] 
\end{equation}   
where $I(E)$ is the spectral energy distribution, $Ag$, is the normalization 
coefficient calculated as;
\begin{equation}
Ag = \frac{E_{total}}{\int\limits_{E_{range}} I(E) \rmn{dE}},
\label{eq:normalization}
\end{equation}
where $E_{total}$ is the total energy output of the \emph{first}
objects within the energy range ($E_{range}$), and the optical depth $\tau (E;r;t)$ is
given by;
\begin{equation}
\tau (E;r;t) = \sum\limits_i \int\limits_r\sigma_i(E) n_i(r;t) \rmn{d}r
\end{equation}
Where the sum is over all species. The effects of secondary
ionizations -- due to the kinetic energy carried by ejected electron
-- have been folded in through the terms $f_H$ and $f_{He}$. These two
terms depend of the ionization state of the medium and given by
\citep{shull};
\begin{equation}
f_H = 0.3908(1-x_{ion}^{.4092})^{1.7592} 
\end{equation}
and,
\begin{equation}
f_{He} = 0.0554(1-x_{ion}^{.4614})^{1.6660}. 
\end{equation}

Here $x_{ion}$ is the ionized fraction of hydrogen. For
ionization, this effect becomes significant when dealing with very
high energy photons ( $>$ 1KeV) and when the medium is neutral to
partially ionized.

Note here that the causality aspect of the code
comes in through the ionization terms. Because, as will be discussed
later, we are presenting a grid code in which the memory of the
ionization history of all the cells prior to the cell under
consideration is embedded in the optical depth, which in turn dictates
the amount of radiation available to ionize a particular cell.

Fits for the collisional ionization and recombination coefficients
were obtained from the appendix of \cite{fuku}. The photoionization
cross-sections though are obtained using the fitting formula of
\cite{verner}. For the exact form of these fitting function we refer
the reader to these two papers.

The temperature evolution is monitored by coupling equations \ref{xhfrac} to 
 \ref{xhe3frac} with the equation below.

\begin{eqnarray}\label{eq:temp}
\lefteqn{\frac{3}{2}\frac{\mathrm{d}}{\mathrm{d}t}\left(\frac{k\rmn{T}_e n_B}{\mu}\right)
=} \nonumber \\
& &  {}  f_{Heat}\hspace{-.5cm}\sum\limits_{\rmn{i=H_I,He_I,He_{II}}} \hspace{-.5cm} n(i) \int \sigma_i (E-E_i) N(E;r;t)~ \frac{\mathrm{d}E}{E} {}  \nonumber \\
& &  {}+ \frac{\sigma_s n_e}{m_e c^2} \sum\limits_{\rmn{i=H_I,He_I,He_{II}}}  \int  N(E;r;t) (E - 4k_B \rmn{T})~ \mathrm{d}E {} \nonumber\\
& &  {}- \hspace{.85cm}\sum\limits_{\rmn{i=H_I,He_I,He_{II}}} \hspace{-.5cm} \zeta_i n_e n(i) \nonumber\\
& &  {}- \hspace{.7cm}\sum\limits_{\rmn{i=H_{II},He_{II},He_{III}}} \hspace{-.5cm}\eta_i n_e n(i) \nonumber\\
& &  {}- \omega_{He_{II}} n_e n_{\rmn{He_{III}}}  \nonumber \\
& &  {}- \hspace{.85cm}\sum\limits_{\rmn{i=H_I,He_I,He_{II}}} \hspace{-.5cm}\psi_i n_e n(i) \nonumber\\ 
& &  {}- \theta_{\rmn{ff}} [n_{\rmn{H_{II}}}+n_{\rmn{He_{II}}}+4n_{\rmn{He_{III}}}] n_e \nonumber \\
& &  {}- 2 \frac{\dot a}{a}\left(\frac{k\rmn{T}_e n_B}{\mu}\right)
\end{eqnarray}

 The form of equation \ref{eq:temp} is identical to that of
\cite{fuku} except for the inclusion of the Compton heating term
(ref. \citealt{madauE}) which become important as we approach the
source which is placed at the centre, i.e., radius equals zero. In the
above equation, $E_i$ is the threshold energy of ionization of the
$i^{th}$ species which is either HI, HeI or HeII. $\sigma_s$, is the
Thompson scattering cross-section of an electron ($\sigma_s = 6.6524
\times 10^{-25} \hspace{.3cm} \rmn{cm}^2$). $\zeta_i$ and $\eta_i$ are
the collisional-ionization cooling and recombination cooling
coefficient, respectively. $\omega_{He_{II}}$ is the dielectronic
recombination cooling coefficient due to $\rmn{He_{II}}$. $\psi_i$ is
the collisional excitation cooling coefficient and $
\theta_{\rmn{ff}}$ the free-free cooling coefficient.  Cooling due to
Hubble expansion is accounted for by the term, $\frac{\dot
a}{a}\left(\frac{k\rmn{T}_e n_B}{\mu}\right)$ where $\frac{\dot
a}{a}\equiv H$ is the Hubble constant. The factor $f_{Heat}$ is
the amount of heat deposited by secondary electrons and is given by;
\begin{eqnarray}
f_{Heat} = \left\{ \begin{array}{ll}
  0.9971(1-(1-x_{ion}^{0.2663})^{1.3163}). &\mbox{ if $x_{ion} > 10^{-4} $} \\
  0.15  &\mbox{ if $x_{ion} \leq 10^{-4} $}
       \end{array} \right.
\nonumber
\end{eqnarray}

 The above expression is an extrapolation of fitting formula used in
\citet{shull} and the results of their calculations as plotted in their Figure
3. We here use the fact that the heating fraction never really goes to zero
but saturates around a value of 0.15. The fitting formulas used in 
\citet{shull} are appropriate for relatively high energies (typically $>$ 
100 eV).

\begin{equation}
f_H(x_{ion},E) = 0.3908(1-x_{ion}^{.4092a(x_{ion},E)})^{1.7592} ,
\label{eq:dijkstra}
\end{equation}
where $a(x_{ion},E)$ is,
\begin{equation}
a(x_{ion},E) = \frac{2}{\pi} \rmn{arctan}\left[ \left( \frac{\rmn{E}}{0.12 \rmn{KeV}} \right) \left( \frac{0.03}{x_{ion}^{1.5}}+1\right)^{0.25} \right].
\end{equation}

Notice that the inclusion of an energy dependent fitting
formula for the fraction of energy that goes to ionization, as in  equation
\ref{eq:dijkstra} \citep{dijkstra1}, does not alter the result
significantly. This is due to the fact that the lower energy photons
are trapped close to the vicinity of source for ionization and the
remaining photons are of relatively high energies for which the
fitting formula of \citet{shull} holds.

The integrals in equation \ref{xhfrac} to \ref{eq:temp} are
pre-calculated and tabulated as a function of $n_ix_i$,
where the index 'i' refers to HI, HeI or HeII. $n_i$ refers to the
abundances and $x_i$ the ionized fraction of the $i$th species.
In other words we are tabulating the integrals as a function of 
optical depth per unit distance. Therefore in principle
we should have a three dimensional table, as is commonly used
\citep{crtcp}. But the integration table in our implementation is at
most two dimensional. We get around using three dimensional tables by
utilizing the fact that the functional form for the photoionization
cross-sections of $\rmn{H_I}$ and $\rmn{He_{II}}$ are the same expect
for a constant difference \citep[see equations (B13) and (B16)
of][]{fuku}.

\subsection{The Algorithm}

The aim of the radiative transport code developed is to compute
the fraction of the ionization for the species of hydrogen and helium
and the temperature for points along a radial direction away from
the source, at various times. This was achieved by solving the time
dependent rate equations (equations~\ref{xhfrac}-\ref{eq:temp}) using
the ODEINT routine of Numerical Recipes \citep{press}, using the
implicit scheme of integration for \emph{stiff} equations. The entire
code was developed in ANSI C.

\begin{figure}
\centering
\hspace{0cm}
\includegraphics[width=.5\textwidth]{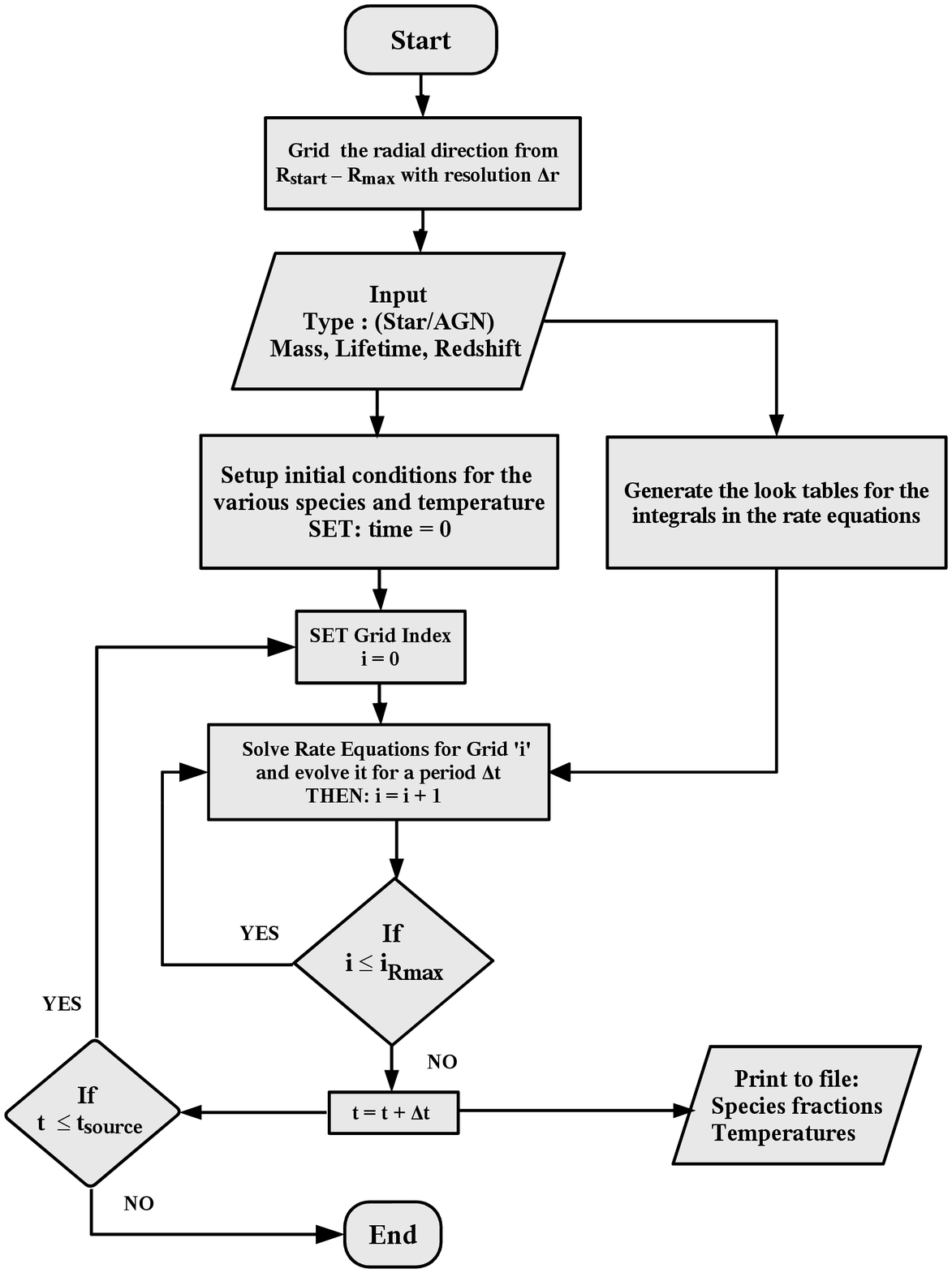}
\vspace{.2cm}
\caption{Flowchart outlining the basic modules in the algorithm 
implementing the radiative transfer code.}
\label{fig:algorithm_flow}
\end{figure}

Figure \ref{fig:algorithm_flow} shows the basic steps involved in the
solution of these rate equations. Firstly, the direction radial to the
source is gridded into cells, the sizes of which are decided as
described later. The inputs to the code at this stage are the parameter 
space we desire to probe in this study which includes the spectrum, masses
of the sources and the redshifts at which these sources switch on. The 
spectrum of the sources, broadly speaking are classified into power-law 
and blackbody. The slopes of the power law and the temperature of the 
blackbody can be set to any desired value. Once the spectrum of the source
is known, a lookup table for the values of the integrals appearing in the
rate equation is created.

Before the start of the simulation, the initial conditions are setup
as follows. The region around the source is set to be completely
neutral with the primordial abundance fraction of hydrogen and
helium. All species therefore have densities corresponding to that of
the IGM at that epoch, $n_\rmn{H}(z)=n_\rmn{H}(0)(1+z)^3$,
$n_\rmn{He}(z)=n_\rmn{He}(0)(1+z)^3$, where $n_\rmn{H}(0)= 1.9 \times
10^{-7}~\rmn{cm}^{-3}$ and $n_\rmn{He}(0)= 1.5 \times
10^{-8}~\rmn{cm}^{-3}$ are the IGM densities at redshift zero of
hydrogen and helium respectively or the initial density
profile could also be set according to a profile as in equation
\ref{eq:stardenprofile}. Abundances of all singly and doubly ionized
species $n_\rmn{H_{II}},n_\rmn{He_{II}}, n_\rmn{He_{III}}$ and
electron density $n_\rmn{e}$ is set to zero.  Initial kinetic
temperature $T_\rmn{e} = T_\rmn{CMB}(z)(1+z)^2/ (1+250)$.  Here we
have assumed that the kinetic temperature of the gas was coupled to
the CMB temperature ($T_\rmn{CMB}$) till a redshift of 250, after
which the $T_\rmn{CMB}$ continued to fall off as $1/(1+z)$ and $T_e$
like $1/(1+z)^2$.

We start our simulation at $\rmn R_{start}$, typically
$0.1$~physical $\rmn{kpc}$ from the location of the source. All
hydrogen and helium are assumed to be completely ionized inside this
radius $\rmn R_{start}$. The outcome of the calculations for different
choices of $R_{start}$ has been tested and, for the choice made in
this paper, found to converge to the required accuracy.

Each cell is then updated for time $\Delta t$, which again
is chosen based on a convergence criterion. This $\Delta t$ is not the
intrinsic time-step used to solve the differential equation itself
because that is adaptive in nature and varies according to the
tolerance limit set in the ODE solver.  On the other hand, the $\Delta
t$ here decides for how long a particular cell should evolve before
moving on to the next. For example, we cannot evolve the first cell
for the entire lifetime of the source and move on to the next
cell. Therefore, $\Delta t$ is decided by reducing it by half each
time until two consecutive runs with $\Delta t$ and $\Delta t/2$ give
the same final result for the ionized sphere within a tolerance
limit. As eluded to before, the code is causal in the sense that a
cell $i$+1 is updated after cell $i$. Note that the light travel time
is not taken into consideration explicitly since the ionization front
(I-front) is very subluminal. Hence, all cells are updated for time $n
\Delta t$ at the $n^{th}$ time-step.

After all cells until the last cell $i_{Rmax}$ are updated for time
$\Delta t$, the resulting values are stored and then passed on as
initial conditions for the evolution of the cell in the next $\Delta t$
interval of time. The update of all $n_{cell}$ is repeated ~$n$~ times
such that ~$n\Delta t = t_\rmn{quasar}$, where $t_\rmn{quasar}$ is the
life time of the miniquasar. The various quantities of interests can
be stored in a file at intervals of choice.

A radial coverage of $R_{max}$ is chosen \textit{a priori} which,
depending on the problem, can be set to any value. Typically we do not
need to go above ten comoving mega-parsecs. This radius is then gridded
equally with a resolution of $\Delta r$, which like the time
resolution, is decreased to half its value until it meets a given
convergence criterion, which here is that the final position of the
I-front is accurate within $0.5\%$. For a typical run, which normally
takes about 100 time steps, this criterion gives an accuracy of about
$5\times 10^{-5}$ per time step.

The block computing the solution of the rate equations in Figure 
\ref{fig:algorithm_flow} requires further explanation. Once the spatial 
resolution is set by following the procedure described above, we
compute for every cell, $\sum_1^{j-1} (n_ix_i)_s$, where $j$ is the
grid-cell under consideration, and these computed sums are used to
evaluate the values of the integrals in equation \ref{xhfrac} to
\ref{eq:temp} from pre-computed tables through a simple polynomial
interpolation. The initial conditions and the value of the integrals
are passed to the solver ODEINT \citep{press} with the \emph{driver}
to solve the \emph{stiff} equations. At the end of this run all cells
are updated for a time $\Delta t$.

\section{Testing the Code}
\label{tests}
 In order to examine the performance of the code we carry out three
tests. The first test is against the analytical form of the
velocity of the I-front with constant background density and central
source emitting a fixed number of photons, namely, the Str\"omgen
sphere case. The second, is to compute the I-front velocity around a
source again with a constant photon flux but with the background 
density evolving with redshift. This is then compared with the
analytical solution of \citet{shapiro}. In both these cases
the I-front is defined as the position at which the ionized and neutral
hydrogen fractions are equal. The third test was a comparison between an  analytical model of heating \citep{zaroubi07} and the RT code. The results agree well except in the discrepancy are discussed in the Appendix.

\subsection{Test-1: I-front velocity in simple cosmology}
The first problem was to test the expanding HII bubble around a source
that produces a fixed number of photons per unit time.  Analytically,
the position $r_\rmn{I}$ and velocity $v_\rmn{I}$ of the I-front can
be written as:
\begin{equation}
r_\rmn{I} = r_\rmn{s}[1-exp(-t/t_\rmn{rec})]^{1/3},
\end{equation}

\begin{equation}
v_\rmn{I} = \frac{r_\rmn{s}}{3t_\rmn{rec}}\frac{exp(-t/t_\rmn{rec})}{[1-exp(-t/t_\rmn{rec})]^{2/3}}.
\label{eq:rI}
\end{equation}
Where $r_\rmn{s}$ is the Str$\ddot{o}$mgen radius \citep{dopita};
$r_\rmn{s} = \left[ \frac{3\dot{N}_\gamma}{4\pi\alpha_B(T)Cn_H^2}
\right ]^{1/3}$; $t_\rmn{rec} = [C\alpha_B(T)n_\rmn{H}]^{-1}$, is the
recombination timescale; $n_\rmn{H}$, the neutral hydrogen density;
$C$ is the clumping factor and $\alpha_B(T) = 2.6 \times
10^{-13}(T/10^4)^{-0.85}~cm^3s^{-1}$ is the hydrogen recombination
coefficient at temperature $T$. The parameters used for this example
are: $\dot{N}_\gamma = 10^{54}~S^{-1}$, $C = 5,$ $\alpha_B = 2.6
\times 10^{-13}~cm^3s^{-1}$ and $n_\rmn{H} = 1.87 \times 10^{-4}~cm^{-3}$.

\begin{figure}
\centering
\hspace{-.6cm}
\includegraphics[width=.5\textwidth]{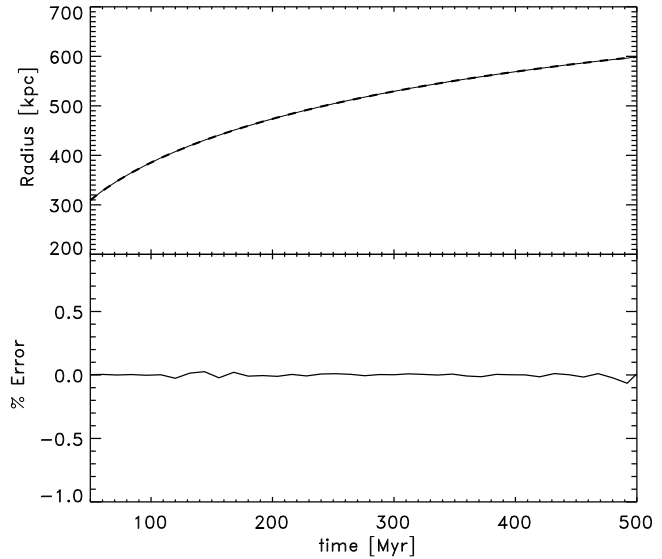}
\caption{Top panel shows the simulated (solid) and the theoretical
(dashed) results of the evolving I-front in an uniform
background. This simulation include only hydrogen of constant density
(i.e., no expansion). }
\label{fig:test1}
\end{figure}

The solid line in Figure \ref{fig:test1} is plotted based on equation
\ref{eq:rI} and the dashed line represents the numerical solution to
the problem. The numerical solution obtained by the RD code is within
$0.1\%$ of the theoretical value.

\subsection{Test-II: The Shapiro \& Giroux test: Ionization 
             front in an expanding Universe}

Shapiro \& Giroux (1987) derived an analytical solution for the
position of the I-front as a function of cosmic time in an expanding
Universe that contains only hydrogen. The solution is of the form,
\begin{equation}
y(t) = \lambda e^{\lambda t_i/t} [t/t_i E_2(\lambda t_i/t)-E_2(\lambda)],
\end{equation}
where $y(t) \equiv (r_I(t)/r_{s,i})^3$, $t_i$ is the age of the
Universe at the time the source has turned on and $r_I$ and $r_{s,i}$
are the comoving I-front position radially from the centre of the
source and the initial Str$\ddot{o}$mgen radius
$(3\dot{N}_\gamma/C\alpha_B(T)n_H^2)^{1/3}$, respectively. Here $n_H$
is the comoving density of hydrogen at source turn-on time.
$\lambda \equiv t_iC\alpha_B(T)n_H$, is basically the ratio between
cosmic source turn-on time and the recombination timescale. $E_2(x)$
is an exponential integral of the form $E_2(x) = \int_1^\infty
{e^{-xt}}/{t^2} dt$.

\begin{figure}
\centering
\hspace{-.6cm}
\includegraphics[width=.5\textwidth]{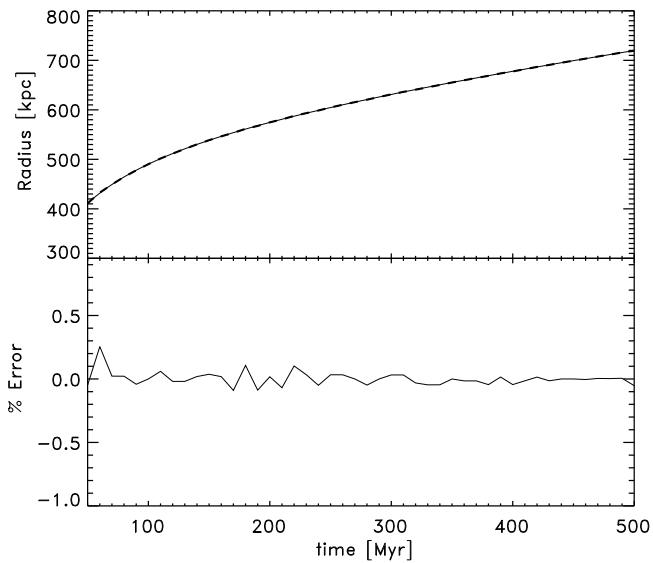}
\caption{Top panel shows the simulated (solid) and the theoretical(dashed)
  results of the evolving I-front in an expanding Universe with background
density being that of the IGM at that epoch.}
\label{fig:test2}
\end{figure}

Result of the test is plotted in Figure \ref{fig:test2}. The numerical
and analytical solutions here also agree to within a couple of tenth of
a percent throughout the evolution of the I-front.

\section{Application to Power-Law Sources (Miniqsos)}
\label{miniqso}
Sources of ionizing radiation with a power-law spectral energy
distribution has been considered by many authors (e.g.,
\citet{madau,kuhlen,fuku,nusser05,zaroubi05,zaroubi07}). The radiative
transport code is applied to these power-law sources where the
ionization and heating patterns around them are studied as a function
of their spectral index, mass and redshifts. For some of the plots in this section we have assumed masses of quasars in the order of $10^6 M_\odot$. Although recent works like \cite{volon07} does show that there could be relatively massive blackholes at early redshifts, a $10^6 M_\odot$ would be unlikely. But we have included this case in our discussion for a completness in spanning the parametre space and also to clearly distinguish the various effects of these sources on the IGM. In the following subsections we introduce and discuss our main findings.

\subsection{Energy spectrum the miniquasar}

 Recent observations and catalogues published in the literature
\citep{vanden,vignali,laor,elvis}, suggest that the energy spectrum
of quasars follows a power-law of the form
$E^{-\alpha}$. Specifically, we explore two types of power law
spectra:
\[ \mbox{I(E)} \propto  \mbox{$E^{-\alpha}$ }
\left\{ \begin{array}{l}
      \mbox{if $\;\; 10.4 \mathrm{eV} < E < 1 \mathrm{keV}    ~\mathrm{(LE~ case)}$}   ;\\
             \mbox{if $\;\; 200 \mathrm{eV} < E < 1 \mathrm{keV}    ~\mathrm{(HE~case)}$}.
\end{array} 
\right. \]
\vspace{-1cm}{ \begin{equation}\label{eq:spectrum}\end{equation}}
\vspace{.3cm} The value of $\alpha$ is fixed to unity for most of the
study, although in section\S\ref{alphavary} we do briefly discuss the
effects of varying this parameter. LE and HE stands for the
'Low Energy' and the 'High Energy' lower limits to which the spectral
energy distribution extents.  The HE case is considered in order to
take into account, in an approximate manner, the possibility of the
lower energy photons being absorbed in the close vicinity of the
source. Instead of single slope one can also adopt multi-slope
spectral templates \citep{saz} but this is not done in this study.

The miniquasars are assumed to accrete at a constant fraction
$\epsilon$ (normally 10\%) of the Eddington rate. Therefore, the miniqso
luminosity is given by:

\begin{eqnarray}
L & = & \epsilon\, L_{\rmn{edd}}(\rmn{M})\\ & = & 1.38 \times 10^{37}
\left(\frac{\epsilon}{0.1}\right) \left(
\frac{\rmn{M}}{\rmn{M_\odot}}\right) \rmn{erg~s^{-1}}.
\label{eq:eddington}
\end{eqnarray}

 The luminosity derived from the equation above is used to normalize
 the relation in \ref{eq:spectrum} according to equation
 \ref{eq:normalization}. The normalization, both for the HE and LE
 case, is done for an energy range of 10.4 $\mathrm{eV}$ to 1
 $\mathrm{keV}$.  Simulations were carried out for a range of masses
 between $10$ and $10^{6} ~\rmn{M}_\odot$. Although the number of
 photons at different energies is a function of the total luminosity
 and spectral index, if we assume that all photons are at the hydrogen
 ionization threshold, then the number of ionizing photons thus
 obtained for the mass range given above is in the order of
 $10^{50}~\rmn{to}~10^{55}$. These are of the same order of magnitude
 of the number of ionizing photons being employed for simulations by
 various authors, \cite{mellema}, \cite{kuhlen}, for example.

\subsection{Radial profile of species' fraction}

\begin{figure}
\centering
\hspace{0cm}
\includegraphics[width=.5\textwidth]{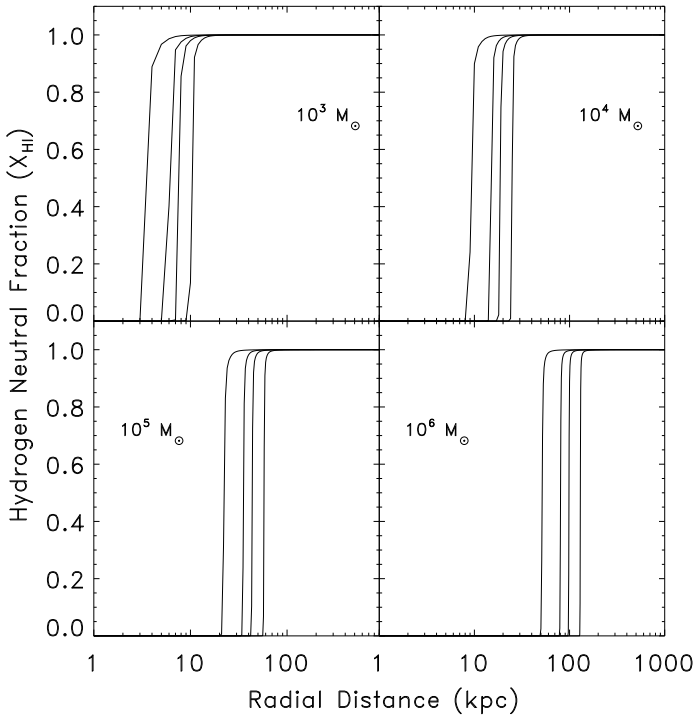}
\vspace{.6cm}
\caption{The neutral hydrogen fraction as a function of distance from the 
 centre of the black hole with masses $10^{3,4,5,6}M_\odot$ (top-left to bottom right panel) at a redshift 
$z=20$ is shown. Four lines in each panel indicates the position of the I-front
after 1,3,5 and 10Myr. These are miniquasars with no UV ionization photons. 
($200 \mathrm{eV}< E <10^4\mathrm{eV}$).}
\label{fig:frach1}
\end{figure}

\begin{figure}
\centering
\hspace{-.3cm}
\includegraphics[width=.5\textwidth]{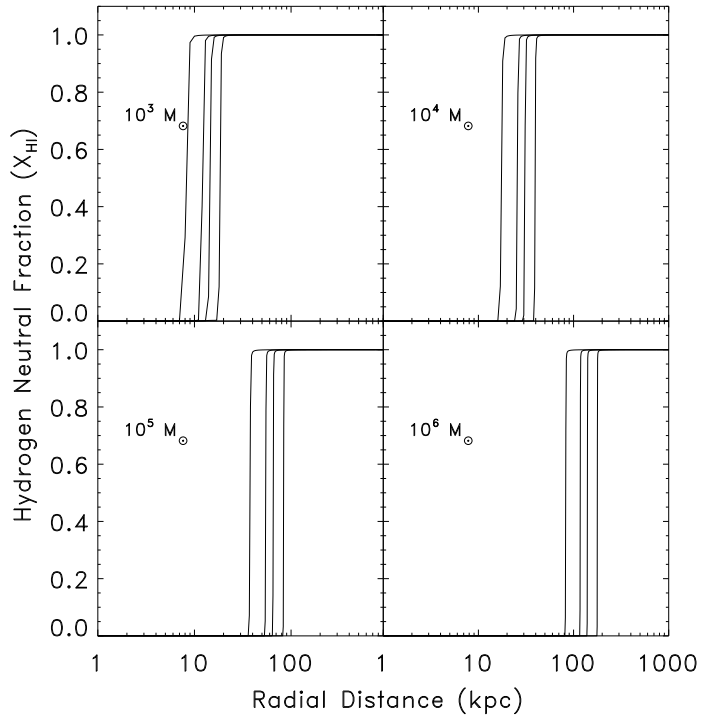}
\vspace{.2cm}
\caption{Same as in fig.\ref{fig:frach1} but now including UV ionizing photons 
($10.4 \mathrm{eV}< E <10^4\mathrm{eV}$).}
\label{fig:LEfrach1}
\end{figure}

Plotted in Figures \ref{fig:frach1} and \ref{fig:LEfrach1} are
snapshots of the neutral fraction as a function of the radial distance
away from the centre of the black-hole for two different cases. The
first (Fig.\ref{fig:frach1}) is the HE case, i.e., without low energy
photons. Whereas, the second (Fig.\ref{fig:LEfrach1}) is for the LE
case , i.e., the photon energy range spans 10.4 eV to $10^4$ eV.  As
indicated on the Figure, each panel corresponds to a different
mass. The lines in each panel corresponds to snapshots of the
neutral fraction at 1, 3, 5 and 10 Myr.  In both cases the sources 
are turned on at a redshift of 20. 
 
Figure~\ref{fig:LEfrach1} clearly shows that the size of the ionized
regions are substantially larger than those in
Figure~\ref{fig:frach1}. Even the 1000 solar mass miniqso produce
regions that are tens of kpc in physical size. This is due to the
increase in the number of photons in the lower energy (LE) range (13.6 to
200 eV), which increases the probability of a photon - atom
interaction relative to the HE case ($\sigma\propto E^{-3}$).
Equilibrium is only reached when the mini-quasar life-time approaches
the recombination time-scale. Since the problem is set in an expanding
Universe we never actually reach an equilibrium solution
\citep{shapiro}. We see that the ionized bubble gets to a couple of
hundred kpc for the case of high energy photons and to more than 0.5
Mpc for $10^6 M_\odot$ miniqsos with low energy photons. All distances
plotted in the Figures are physical unless specified otherwise.

%
%

%
%

\begin{figure}
\centering
\hspace{-.5cm}
\includegraphics[width=.5\textwidth]{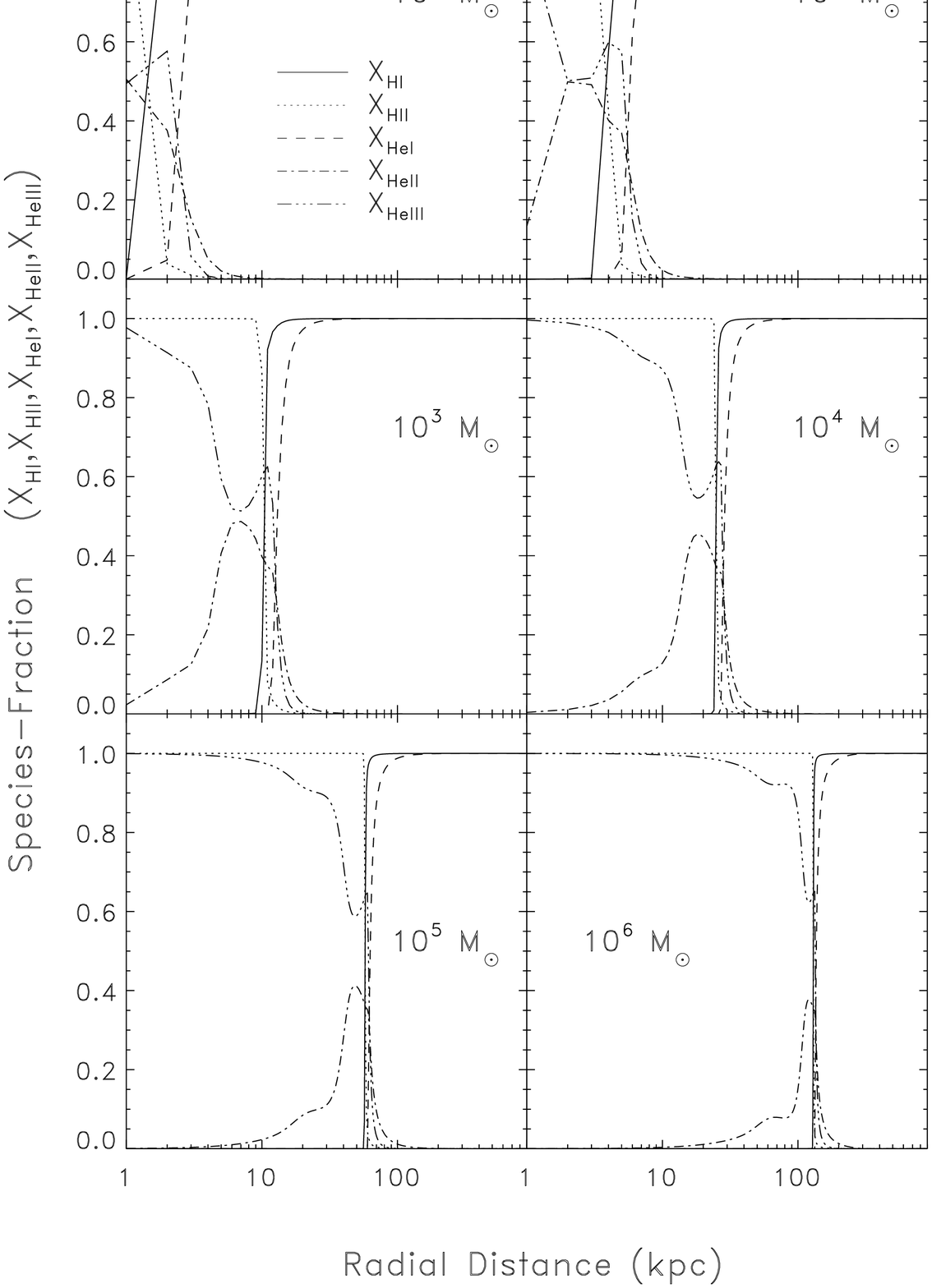}
\vspace{.3cm}
\caption{Fraction of various species
($x_\rmn{HI}$,$x_\rmn{HII}$,$x_\rmn{HeI}$,$x_\rmn{HeII}$, $x_\rmn{HeIII}$ ) as function of
distance from the centre of the black hole with masses $10^{1,2,\dots
6}M_\odot$ (top-left to bottom right panel) at a redshift
$z=20$. These are snapshots after 10 Myr of evolution. Miniquasars
have no UV ionization energy ($200 \mathrm{eV}< E <10^4\mathrm{eV}$).}
\label{fig:fracmix}
\end{figure}

\begin{figure}
\centering
\hspace{-.5cm}
\includegraphics[width=.5\textwidth]{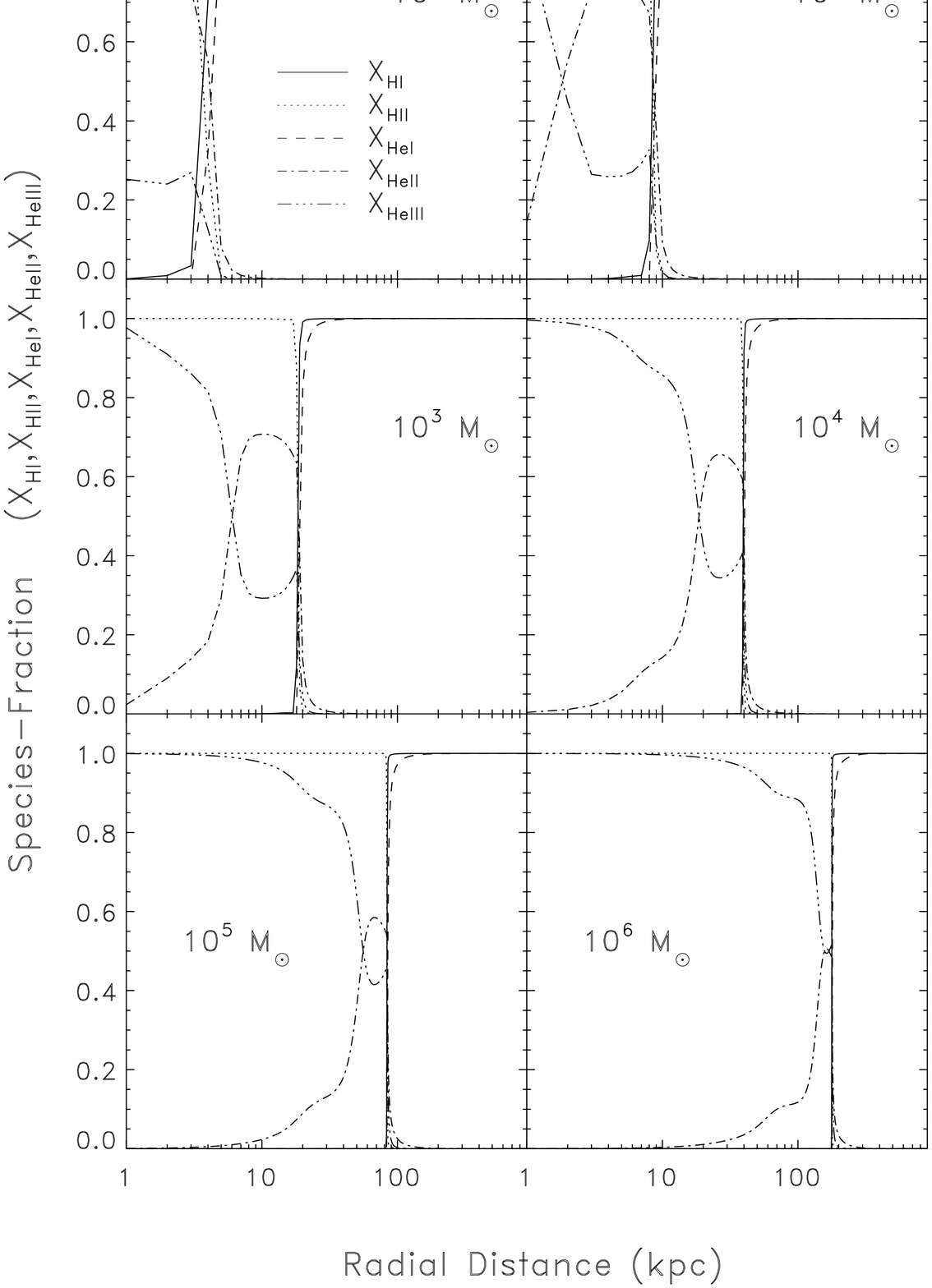}
\vspace{.3cm}
\caption{Same as in fig.\ref{fig:fracmix} but now including UV ionizing photons
  ($10.4 \mathrm{eV}< E <10^4\mathrm{eV}$).}
\label{fig:LEfracmix}
\end{figure}

The abundance of HI, HII, HeI, HeII and HeIII as a function of radial
distance are tracked in time. Figures~\ref{fig:fracmix} and
\ref{fig:LEfracmix} shows the fraction of these species after 10 Myr
of evolution as a function of radius. The miniquasar is switched
on at redshift of 20.

The interplay between the evolution of these species among each other
and with the gas temperature provides us with rich and interesting
structures in profiles. Miniquasars of 10 and 100 solar masses are not
able to produce substantially high number of ionizing photons and
hence the ionized regions around them are relatively small. However,
miniquasars of higher masses like $10^5$ and $10^6$ solar masses
produce ionized bubbles of a couple of comoving megaparsecs.

An interesting detail apparent in these Figures is that the HeI
 I-front exceeds that of the HI. Explanation for this lies in the fact
 that we have relatively high energy photons ($ E > 50~eV$) and the
 cross-section of HI is much smaller than HeI,
 $\sigma_{HeI}(E)/\sigma_{HI}(E) \approx 20$ at an energy E =
 50 eV. Therefore, the probability of high energy photons being capture
 by HeI is higher, increasing the helium I-front distance from the
 centre.

Results are shown for the HE (\ref{fig:fracmix}) and LE
(\ref{fig:LEfracmix}) cases. As expected the ionized fronts of
HI and HeI have travelled a much greater distance in the
LE case. This is consistent because the HE spectrum is not only devoid
of hydrogen ionizing photons but also lacks helium ionizing photons
(54.4 eV).
%
%

%
%

Another point to notice, is the visible excess of HI fraction in the
HE spectrum just before the full ionization front in
Figure~\ref{fig:frach1}, especially at the 10 Myrs curves. We
interpret this excess as a result of the interplay between the
increase of HeII and decrease of HeIII as seen in
Figure~\ref{fig:fracmix} thus increasing the HI recombination rate on
the one hand and decreasing photon flux as a function of radius
on the other. This interpretation is supported by the weakening of the
HI excess feature as a function of black hole mass. The phenomena is
not manifested in the LE spectrum (Figure~\ref{fig:LEfrach1}) due to
the ionization efficiency of the UV photons.

\subsection{The kinetic temperature profile}

The heating and cooling terms included in equation \ref{eq:temp} are
coupled with the rate equations \ref{xhfrac} to \ref{xhe3frac}. The
principal heating terms are the bound free and Compton
heating. Bound free heating is by far the most dominant, although
towards the centre Compton heating does become important.

\begin{figure}
\centering
\hspace{1cm}
\includegraphics[width=.5\textwidth]{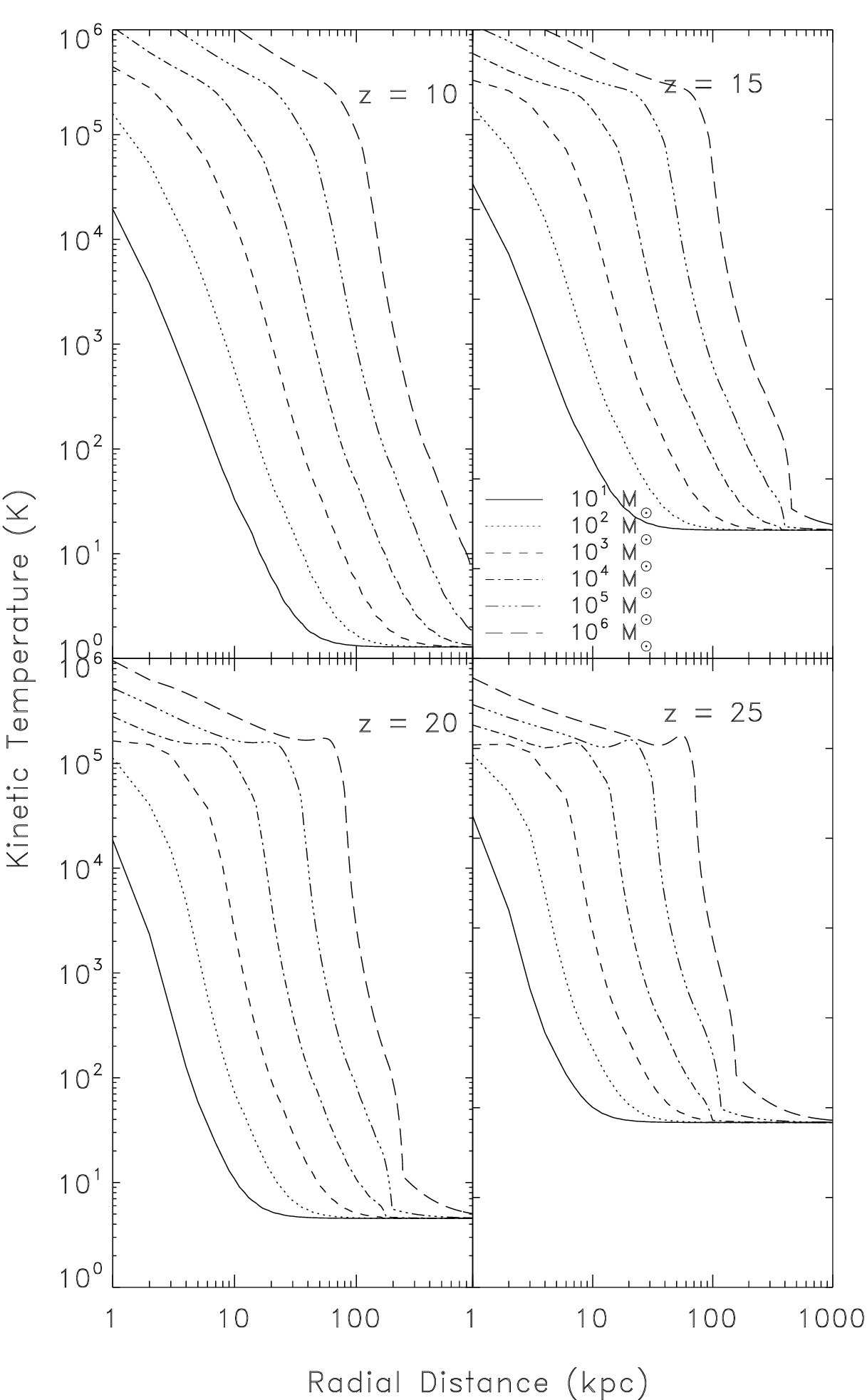}
\vspace{.4cm}
\caption{Radial profile of Kinetic temperature $T_e$ for black hole
with masses $10^{1,2,\dots 6}M_\odot$ at 4 different redshift
$z=10, 15, 20, 25$ (top-left to bottom right panel) after 3 Myrs
is shown. The spectrum of these quasars include only high energies. }
\label{fig:tkin}
\end{figure}

\begin{figure}
\centering
\hspace{10cm}
\includegraphics[width=.5\textwidth]{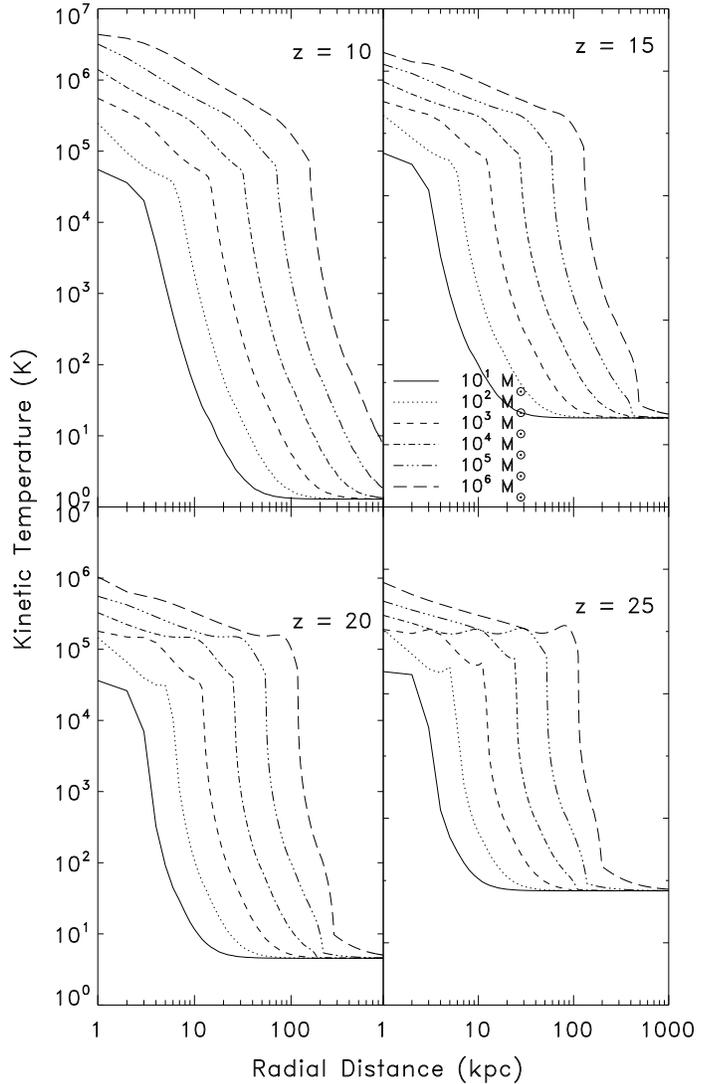}
\vspace{.4cm}
\caption{Same as in fig.\ref{fig:tkin} but these quasars do include the low energy photon. }
\label{fig:LEtkin}
\end{figure}

Figures \ref{fig:tkin} and \ref{fig:LEtkin}, corresponding to the HE
and LE cases respectively, show the temperature profile for masses
from $10 M_\odot$ to $ 10^6 M_\odot$ for four different redshifts. The
snapshot is taken after 3 Myrs of evolution.

Although there is a substantial change in the ionization profiles
(sizes of the ionized spheres), the heating remains more or less the
same except for a few details. The flux of high energy photons, those
responsible for the dominant secondary heating is abundant causing a
sort of invariance in the extent of heating in both cases. Miniqsos of
high masses do maintain an extended volume of high temperatures, in
some cases up to 5 comoving Mpcs.  If this indeed is the case at
relatively low redshifts like around six, then they will be observable
in the Sunyaev-Ze'ldovich effect as discussed in section
\S\ref{szsection}

The four panels corresponding to different redshifts do not show a
significant difference in the heating or its extent. Whereas the
ionized bubbles (Fig.  \ref{fig:frach1} and/or \ref{fig:LEfrach1}) are
strongly dependent on the redshift. This is due to the fact that lower
energy photons (responsible for ionizing) see a much higher optical
depth in neutral hydrogen than do high energy photons. Thus an
increase in density by an order of magnitude (from z=10 to z=25) does
not alter the heating considerably.

%
%

%
%

\subsection{Influence of different spectral energy distributions}
\label{alphavary}
\begin{figure}
\centering
\hspace{-.75cm}
\includegraphics[width=.5\textwidth]{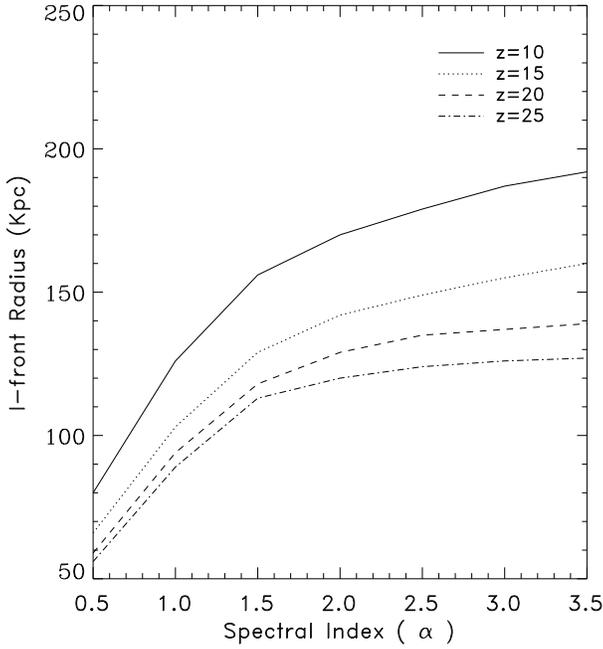}
\vspace{.1cm}
\caption{Position of the I-front is plotted in physical size as a
function of spectral index $\alpha$, 10 Myr after the quasar was
switched on . Note that the sizes of the ionized region plateaus
after about $\alpha=2.5 $. The black hole in the centre is of $10^5
M_\odot$. }
\label{fig:alpha}
\end{figure}

\begin{figure}
\centering
\hspace{-.75cm}
\includegraphics[width=.5\textwidth]{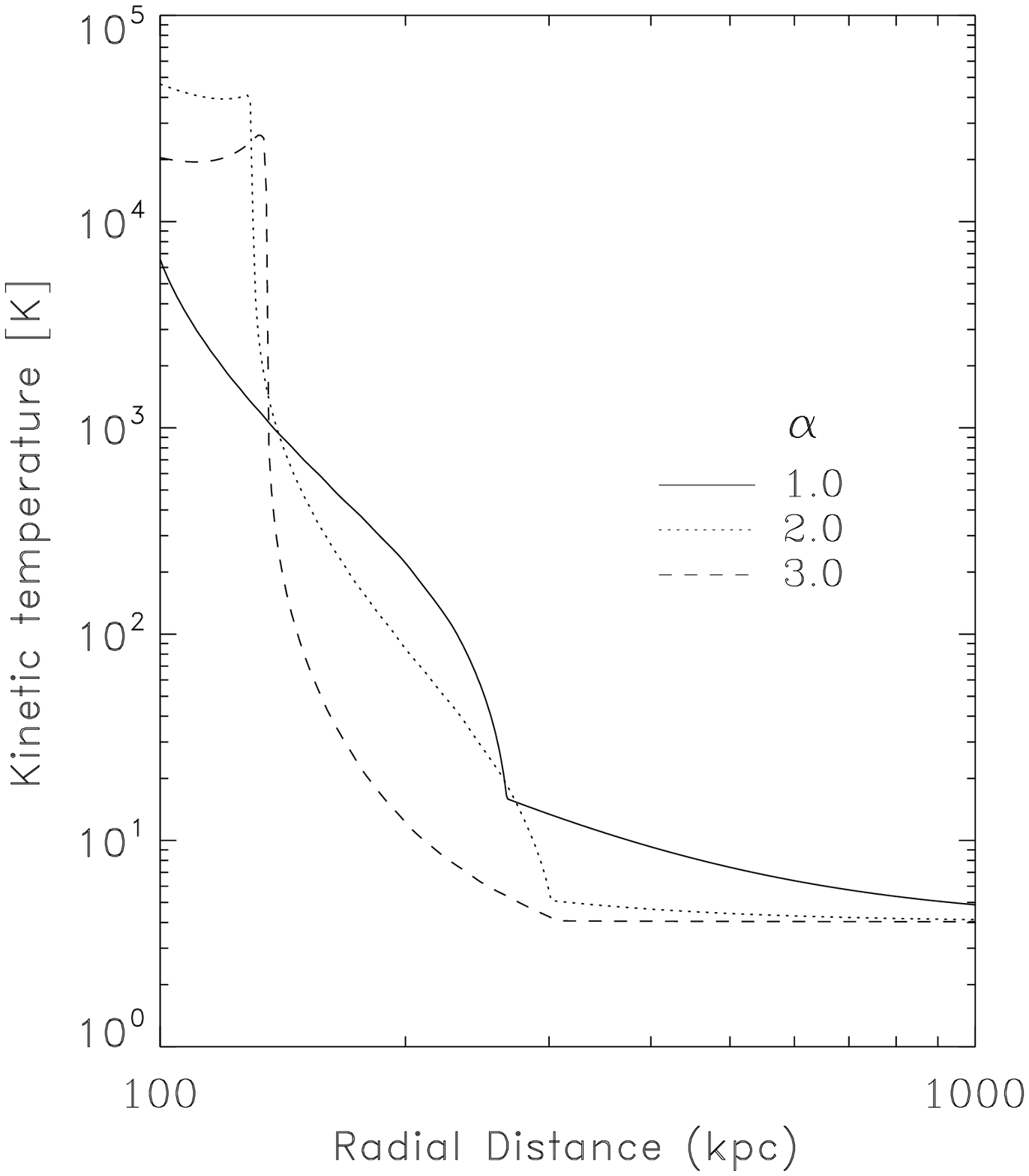}
\vspace{.1cm}
\caption{Temperature around a $10^6 M_\odot$ black hole is plotted for
three different spectral indices at z=20. We see the decrease in
heating with steeper indices. Decrease of low energy photons being the
cause of this behaviour.  }
\label{fig:alphatemp}
\end{figure}

Throughout our study in this paper we have dealt with $\alpha = 1$
case for equation \ref{eq:spectrum}. Here we explore the possibility
of other indices for the power law. As an example we plot
(ref. Fig. ~\ref{fig:alpha}) the size of the ionized bubbles after 10
Myr as a function of $\alpha$ around a quasar with central black hole
of $10^5 M_\odot$ for four different redshifts as indicated in the
figure. Figure \ref{fig:alpha} shows that the size of the ionized
bubble\footnote{We define the size of the ionized bubble as the position
of the I-front when the ionized fraction is 0.5 for hydrogen} increases
with $\alpha$. This is expected because an increase in $\alpha$ for
the same given normalization concentrates more photons at the low
energy end, i.e., around 13.6 eV and fewer high energy photons. Thus,
for higher $\alpha$, one expects more ionization and less
heating. This indeed is reflected in Figure~\ref{fig:alphatemp}.

The central regions are heated more or less to the same extent. In
order to see the significant difference we have only plotted the
distances between 100 and 1000 kpc physical. It is clearly seen that
for steeper slope heating drops considerably.

%
%
%
%

\subsection{Life-time \& Duty-cycle of mini-quasars} \label{dutycycle}

  \citet{wyithe02} produce analytical estimates of the quasar-life
time and duty cycle (see their Figure~3). Certainly this sort of
periodic switching on and off of the quasar leaves an imprint on the
IGM. We simulate a case wherein the quasar duty cycle is 100~Myr with
an active phase of 10~Myr at the start of the cycle.

\begin{figure}
\centering
\hspace{-.5cm}
\includegraphics[width=.5\textwidth]{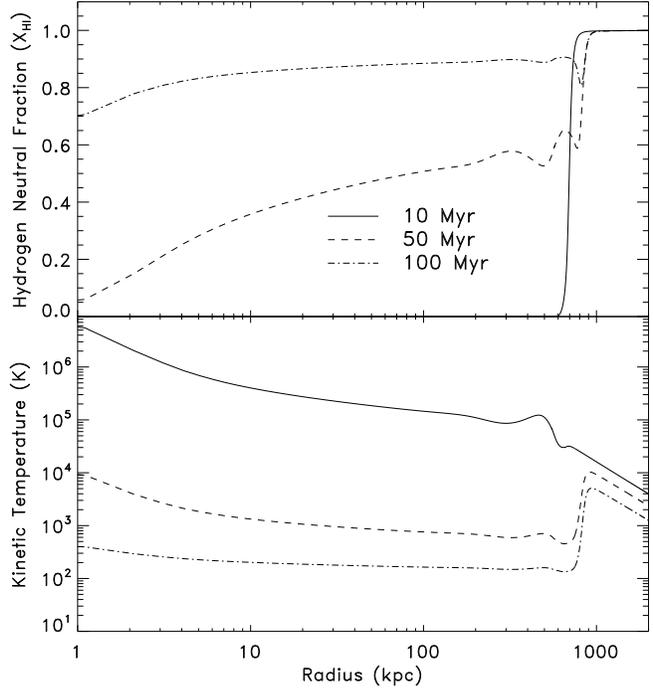}
\vspace{.3cm}
\caption{Figure shows the neutral fraction (top panel) and kinetic
temperature (bottom panel) for three different times after the quasar
was switched on as indicated in the figure. The quasar is at redshift
$z = 10$ and hosts a 1000 $M_\odot$ black hole in the centre, which is
switched on for the first 10 Myr and switched off since then. }
\label{fig:compareONOFF}
\end{figure}

 Recombination time-scales are orders of magnitude larger than the
ionization timescales at the mean densities of the IGM. As a result,
once the IGM is ionized by a source with finite life time (say 10Myr),
it takes considerably long time for the IGM to recombine, leaving
behind a bubble of ionized gas. Figure \ref{fig:compareONOFF} shows
the ionized fraction and the temperature profile just after the source
is switched off at 10 Myr, than at 50 Myr and at the end of the duty
cycle which is 100 Myr. The source is a miniqso of a 1000 $M_\odot$ at
redshift 20.
 
 There are a couple of points to be noted here. One is the fact that the
I-front position is further away at 50 and 100 Myr than at 10 Myr
although the source was switched off after 10 Myr. This propagation of
the I-front in the absence of photons from the source is attributed to
the ionization due to collisions. This continues until the
temperature drops sufficiently so that collisions become
ineffective. In fact before the IGM recombines completely the quasar
is switched on again. Thus we see \footnote{movies
created from the simulation can be obtained by contacting the authors}
that the ionizing front expands during the on-time, stands still
around the same position during the off-time and continues to expand
during the next on-time and so on.

The drop in temperature on the other hand is easily seen within this
turn-off period. Compton cooling is the dominant sink for the
temperature.  Also we see that the regions close to the centre remain
relatively more ionized than the rest. This is again a consequence of
the temperature profile.  Regions with higher temperature continues to
be ionized for longer periods.

%

\begin{figure}
\centering
\hspace{-.5cm}
\includegraphics[width=.5\textwidth]{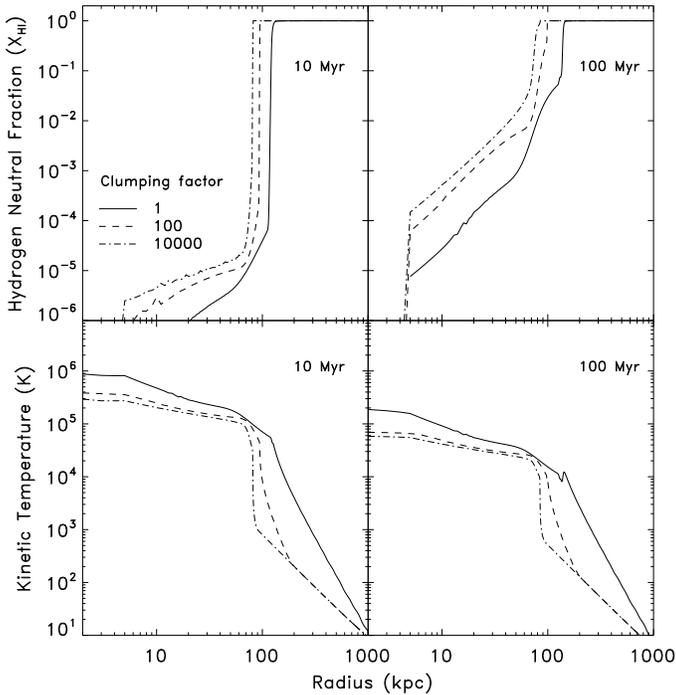}
\vspace{.3cm}
\caption{Figure shows the neutral fraction (top panels) and kinetic
temperature (bottom panels) for three different clumping factors at 10
Myr and 100 Myr as indicated. The quasar is active at redshift $z = 20$
and hosts a 1000 $M_\odot$ black hole in the centre, which is switched
on for the first 10 Myr and switched off since then. }
\label{fig:compareDENS}
\end{figure}

All of the analysis performed thus far was done by embedding the
source in a uniform background. But we know that objects like quasars
are preferentially formed in overdense regions. In order to test the
impact of different overdensities on the I-front position and
temperature produced by a quasar we embedd the quasar in a background
that is 1,10 and 100 times the mean IGM density. The results are shown
in Figure \ref{fig:compareDENS}. The miniqso has a 1000 $M_\odot$
black hole at the centre and is at a redshift of 10 and contains only
the high energy photons. As expected the ionization fronts could
penetrate less at higher densities. But the recombination rate is much
greater at higher densities. This is seen in the top right panel
corresponding to 100 Myr as the I-front corresponding to a clumping
factor of a 100 has recombined much more. On the other hand the
heating is relatively less at higher densities because most of the
photons are absorbed to ionize and the dominant cooling terms, e.g.,
like the Hubble cooling, are enhanced at higher densities.

\section{Application to stellar ~(blackbody) sources}
\label{stars}
Unlike quasars and miniqsos that have a power-law distribution of
energy in their spectrum, the stars approximately behave as a
blackbody of a given temperature. This blackbody nature of the source
leaves different signatures in the manner the IGM is heated and
ionized.  Given the weak dependence of the star's temperature on its
mass (see \citet{schaerer}), we fix the blackbody temperature of the
stars to $5 \times 10^4~\mathrm{K}$ and mass ranges between $10$ and
1000 $M_\odot$. The total luminosity for a given stellar mass is
calculated from table~3 of \citet{schaerer}. The other difference in case of stars is the density profile in which it is embedded. We assumed a density profile of the form;
\begin{eqnarray}
 \rho(r) &=& 3.2 \times (91.5~\mathrm{pc}/r)^2 ~ [\mathrm{cm}^{-3}] ~~~ (r \le  r_{c}) \nonumber \\
          &=& n_i(0) \times (1+z)^3 ~ [\mathrm{cm}^{-3}] ~~~~~ (r >  r_{c}),
\label{eq:stardenprofile}  
\end{eqnarray}
where  $n_i(0)$, is the density of the hydrogen or helium at redshift 0 and
$r_c$ is the radius at which the density profile falls to that of 
the mean IGM.

Depending on the temperature, the spectrum used above for the stars
peaks in between $\approx 20~eV $ to $\approx 24~eV$ (for
$10^5~\mathrm{K}$).  What is expected in the case of stars of the same
power output as a quasar are bigger ionized bubbles and less
heating. Simply because, after the peak, which is around the
ionization threshold of hydrogen and helium, we have an
exponential cutoff towards higher frequencies.

\begin{figure}
\centering
\hspace{-.6cm}
\includegraphics[width=.5\textwidth]{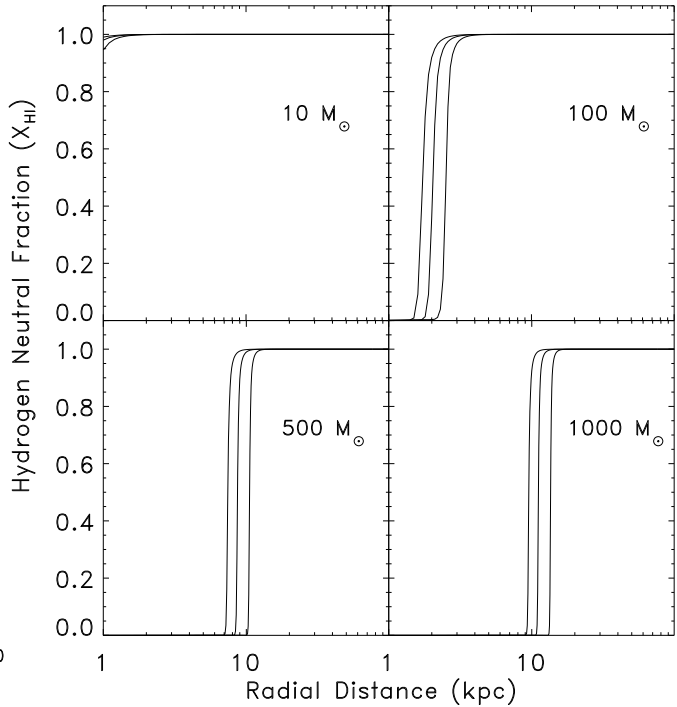}
\vspace{.5cm} 
\caption{The neutral hydrogen fraction as a function of distance from the 
 centre of the blackbody with an effective temperature of 50000K and masses 
 10, 100, 500 , 1000 $M_\odot$ (top-left to bottom right panel) at a redshift 
$z=20$. The three lines corresponds to the I-front position after 2,3 and 5 Myr
 of evolution.}
\label{fig:starfrach1}
\end{figure}

After modifying the spectrum and the underlying density profile, we
basically did the same exercise as for the miniqsos. Some of the
results are plotted in the following Figures.  The hydrogen neutral
fraction is plotted in figure \ref{fig:starfrach1} as a function of
distance from the source with an effective blackbody temperature of
50000 K. Note that unlike the miniqsos case the x-axis here runs only
up to 100 kpc physical. This is because the most massive stars
considered here, i.e., 1000 $M_\odot$, has a luminosity of a 10
$M_\odot$ black hole shining at the Eddington rate \citep{schaerer}.

%
%

\begin{figure}
\centering
\hspace{-.6cm}
\includegraphics[width=.5\textwidth]{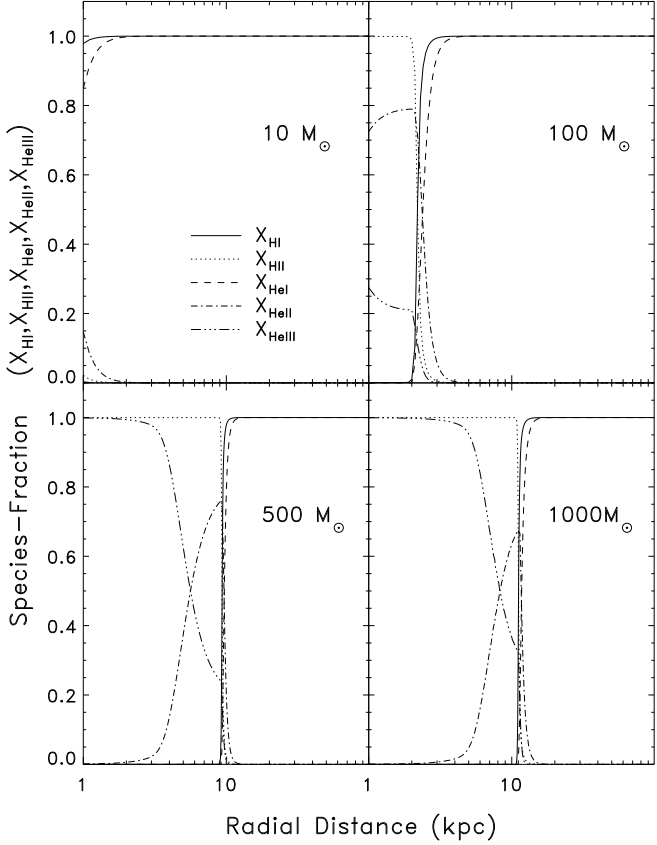}
\vspace{.5cm}
\caption{Fraction of various species ($x_\rmn{HI}$,$x_\rmn{HII}$,$x_\rmn{HeI}$,$x_\rmn{HeII}$,
 $x_\rmn{HeIII}$ ) for the shown star masses.}
\label{fig:starfracmix}
\end{figure}

A similar case of lower ionized helium is seen in Figure \ref{fig:starfracmix}.
Observe again that the helium front is leading. Because for high temperature 
stars the blackbody peaks at higher energies closer to the ionization threshold
of helium.
%
%

\begin{figure}
\centering
\hspace{-.3cm}
\includegraphics[width=.5\textwidth]{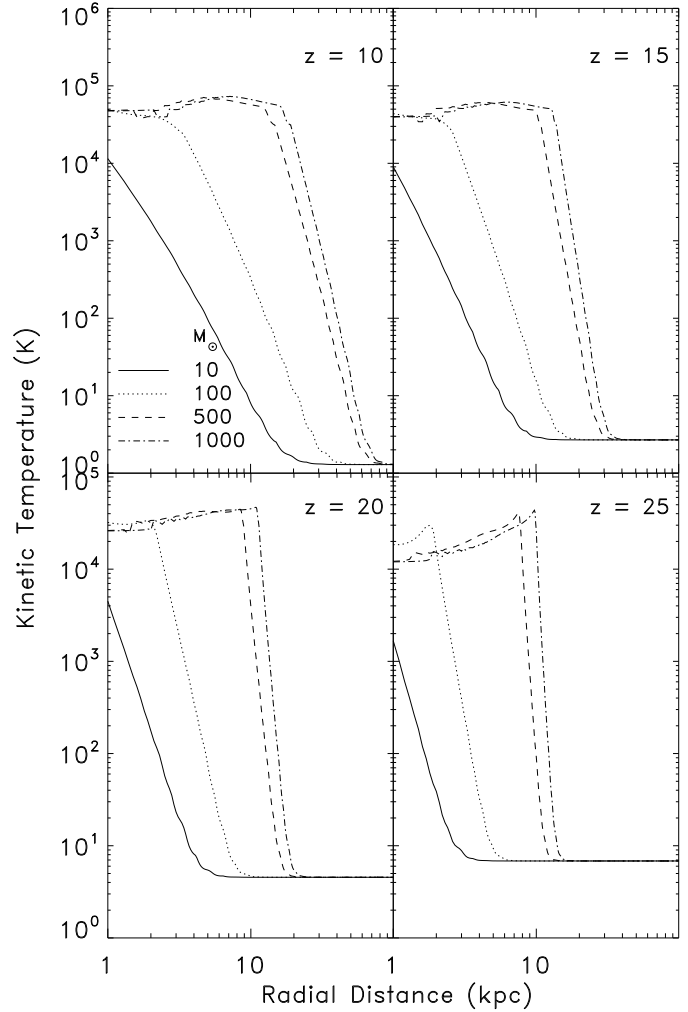}
\vspace{.2cm}
\caption{Radial profile of Kinetic temperature $T_e$ for stars with masses
 $10$, $100$, $500$ and $1000M_\odot$ at 4 different redshift $z=10, 15, 20, 25$ (top-left to bottom right panel) are shown. }.
\label{fig:startkin}
\end{figure}

Figure \ref{fig:startkin} shows kinetic temperature as a function of
radial distance for Population III stars with masses of
$10-1000~M_\odot$. Notice that even though the maximum temperatures
reached are comparable to that of the miniqso, the extent and shape of
the profile are distinctly different. There is a much sharper edge to
the heating attributed to the fact the there are not many  high
energy (100 eV and above) photons and hence the mean free path of most
photons is lower.

%
%

\section{Observational Effects}
\label{observational}

There are many indications of an epoch of reionization in the
Universe. But observations currently are only able to provide us with
either an integral limit (CMB data) or a lower limit (Gunn-Peterson
troughs) on the redshift of reionization. Direct detection and study
of this epoch currently rests on the future 21 cm radio observations
and temperature fluctuations because of the SZ effect (detectable by
PLANCK
\footnote{http://www.rssd.esa.int/index.php?project=PLANCK}) and
probable direct observations of very high redshift Population III
stars by JWST
\footnote{http://www.jwst.nasa.gov/}.

In this paper we will concentrate on the feasibility of radio
observations on the detection and mapping of the epoch of
reionization. A point to note is that from here on in the case of
miniqsos we only use the low energy case because the heating would be
less exaggerated and probably closer to reality.

\subsection{Spin Temperature}

The spin temperature $T_s$ couples to the either the CMB temperature 
$T_\rmn{CMB}$ or to the kinetic temperature $T_e$, in the absence of other 
radio sources \citep{field,kuhlen2006}. Thus $T_s$ can be written as a 
weighted sum between $T_e$ and $T_\rmn{CMB}$,
\begin{equation}
T_s = \frac{T_\star+T_{CMB}+y_{col}T_e+y_\alpha T_e}{1+y_{col}+y_\alpha}.
\end{equation}
$T_\star = h\nu_{21cm}/k = 0.0681~\rmn{K}$, $y_{col}$ and $y_\alpha$
determines the efficiency of collisional and Ly${\balpha}$ coupling
and is given by,
\begin{equation}
y_{col} = \frac{T_\star}{A_{10}T_e} (C_H + C_e +C_p)
\end{equation}
where, $A_{10} = 2.85 \times 10^{-15} ~s^{-1}$ is the spontaneous emission rate
 or the Einstein A-coefficient and $C_H$, $C_e$, and $C_p$ are the 
de-excitation rates of the triplet due to collisions with neutral atoms, 
electrons, and protons, respectively. The empirical fits for these coefficients
 are identical to those used by \citet{kuhlen2006}, which is a combination of 
results published in \citet{zygelman},\citet{allison}, \citet{liszt} and 
\citet{smith}.
And the Ly$\alpha$ coupling coefficient (discussed in section \S\ref{Lysect});
\begin{equation}
y_\alpha = \frac{16\pi^2T_\star e^2f_{12}J_o}{27A_{10}T_em_e c}
\label{eq:yalpha}
\end{equation}

Here, $J_o$ is the Ly$\alpha$ flux density. For the miniqsos high
energy photons Ly$\alpha$ coupling is mainly caused by collisional
excitation due to secondary electrons \citep{chuzhoy06}. This process
is accounted for by the following integral,
\begin{equation}
J_o(r) = \frac{\phi_\alpha c}{4 \pi H(z) \nu_\alpha} n_{HI}(r) \int\limits_{E_o}^\infty \sigma(E) N(E;r) dE, 
\label{eq:lyalphaflux}
\end{equation}
where $\mathrm{f_{12}} = 0.416$ is the oscillator strength of the
Ly$\alpha$ transition, $e$ \& $m_e$ are the electron's charge and
mass, respectively. $\phi_\alpha$ is the fraction of the absorbed
photon energy that goes into excitation \citep{shull}. The contribution of
this term is important close to the miniquasar.

Note that in reality, we have to include the frequency dependence of the
Ly$\alpha$ cross section, the background continuum Ly$\alpha$ photons
and scattering \citep{miralda}. But since we are dealing only with
``injected'' photons at the Ly$\alpha$ frequency we only include the
cross section at the line centre with thermal broadening.

%
%
\begin{figure}
\centering
\includegraphics[width=.5\textwidth]{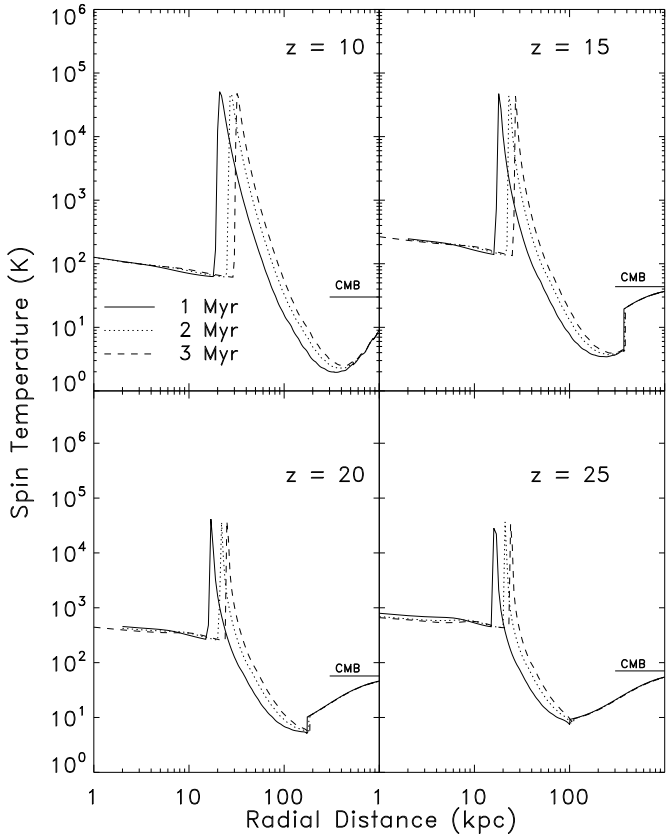}
\vspace{.4cm}
\caption{Radial profile of spin temperature $T_s$ for a black hole
  with mass $10^4M_\odot$ at 4 different redshift $z=10, 15, 20, 25$
  (top-left to bottom right panel) for 1,2 and 3 Myrs of
  evolution after the source was switched on. Note that the curves
  identical to that of the kinetic temperature in Fig.\ref{fig:LEtkin}
  except the spin temperature climbs back to $T_\rmn{CMB}$ after about
  1000 kpc corresponding to the photon propagation distance in 3
  Myrs. These miniquasars include the low energy photons.}
\label{fig:LEtspin}
\end{figure}

Results of the spin temperature for a black hole of mass $10^4M_\odot$ 
is shown in Figure \ref{fig:LEtspin} for 4 different redshifts. For reasons 
discussed in the next section the spin temperature follows the kinetic
temperature for considerable distance away from the source and then 
follows the background CMB temperature at larger distances.

\begin{figure}
\centering
\hspace{-.3cm}
\includegraphics[width=.5\textwidth]{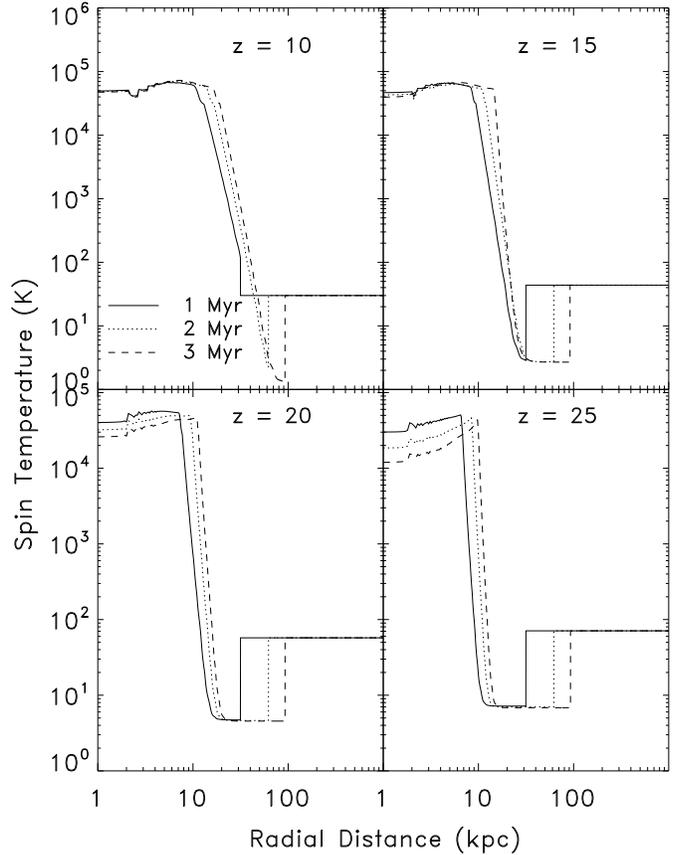}
\vspace{.5cm}
\caption{Radial profile of spin temperature $T_s$ for stars with a mass
 $1000M_\odot$ at 4 different redshifts $z=10, 15.0, 20, 25$ (top-left to bottom right panel) are shown. The temperature of the star is assumed to be 50000K.} 
\label{fig:startspin}
\end{figure}

The spin temperature for the case of a $1000M_\odot$ at 4 different
redshifts is shown in Figure \ref{fig:startspin}. The spin
temperature, for the case of stars, is assumed to follow the kinetic
temperature for a distance of `ct', although this might be an over
estimate because there is a sharp cut-off towards red ward of the
black body peak reducing the number of Ly${\balpha}$ produced by the
source itself. But with a combination of the background Ly${\balpha}$
, it might be possible to couple the spin temperature to the kinetic
temperature.
%
%

\subsection{Lyman ${\balpha}$ coupling}
\label{Lysect}

 Ly$\alpha$ coupling is known to be the most efficient mechanism to
couple the kinetic temperature to the spin temperature, at least in
the mean IGM.  Collisional coupling only becomes important in overdense
regions with gas over-densities $\delta \geq 20[(1+z)/10]^{-2} $
\citep{ilian,fur_l}.  Since we are interested in studying brightness
temperature fluctuation relatively far away from the source, the
dominant mode of coupling for our purposes is Ly$\alpha$ pumping. 

There are a number of ways to generate Ly$\alpha$ photons. 1) The
source itself produces continuum photons between Ly$\alpha$ and
Ly$\beta$, which is then redshifted at a distance from the source to
Ly$\alpha$. 2) Photons above the Ly$\beta$ produced by the source
pumps the electrons to higher levels which then cascade back to
produce Ly$\alpha$. 3) Secondary electrons can deposit a part
of their energy toeards exciting the hydrogen atom \citep{shull}. 4)
If we are looking at sources at lower redshifts ($<20$) then there is
enough background Ly$\alpha$ from ``first stars'' \cite{ciardi2}.

The Ly$\alpha$ flux density estimated in equation \ref{eq:lyalphaflux}
is efficient to couple the spin temperature to the kinetic
temperature. The Wouthuysen-Field effect
\footnote{The Wouthuysen-Field mechanism basically is the mixing of
the hyperfine state in neutral hydrogen in its ground state via
intermediate transitions to the 2\emph{p} and above states
\citep{wouthuysen, field}} is efficient within the photon propagation
(or light travel) time if these photons come directly from the source
within the energy range of ionization (13.6 eV) and Ly$\alpha$ (10.4
eV). But, even in the case when the source does not produce
Ly$\alpha$ photons, the secondary excitations are efficient enough to
couple $T_s$ with $T_e$ to a large distance away from the source as
seen from Figure.\ref{fig:LEtspin}. But at lower redshifts the
background Ly$\alpha$ flux would be large enough to de-couple $T_s$
from $T_\rmn{CMB}$ much further away from the source \citep{ciardi2}.
In the case of stars we do not expect a major influence of the 
secondary electrons on ionizations or excitations. Thus we assume that the
Ly$\alpha$ flux originates from the source/background. Hence, we allow the 
spin temperature to follow the kinetic temperature to the light travel 
distance ``ct''. 

\subsection{Brightness temperature}

Under the ``low-frequency'' or Rayleigh-Jeans' approximation, the equation for 
blackbody radiation reduces to,
\begin{equation}
I(\nu) = \frac{2\nu^2}{c^2}kT_b.
\end{equation}
Where $I(\nu)$ is the intensity of radiation at frequency $\nu$, $k$ the 
Boltzmann constant, $c$ the speed of light and $T_b$ is the brightness 
temperature. And in radio astronomy this limit is applicable. Brightness 
temperature can be measured differentially as a deviation from the background
CMB temperature $T_\rmn{CMB}$ \citep{field,ciardi2} as,
\begin{eqnarray}
\delta T_b & = &26mK~ (1+\delta)~ x_\rmn{HI}\left(1-\frac{T_\rmn{CMB}}{T_s}\right) \left(\frac{\Omega h^2}{0.02}\right) \nonumber \\
        & &      \left [\left(\frac{1+z}{10}\right)\left(\frac{0.3}{\Omega_m}\right)\right]^{1/2}.
\label{eq:dtb}
\end{eqnarray}
Where $x_\rmn{HI}$ is the neutral hydrogen fraction~(see
Fig~\ref{fig:frach1}), $\delta$ is the overdensity of hydrogen (atoms
and ions) and $T_s$ is the spin temperature~(Ref
Fig~\ref{fig:LEtspin}). So given the cosmology and the redshift, which
translates to a frequency of observation, the brightness temperature
in equation \ref{eq:dtb} basically reflects the source characteristics
in two ways. One that enters through the neutral fraction $x_\rmn{HI}$,
defined by the ionizing capabilities of the source and another
through the spin temperature $T_s$ which in effect reflects kinetic
temperature dictated by the heating capacity of the source. Thus, we
hope that a 3D tomography could not only reveal a statistical signal,
the power spectrum and the large scale structure of the epoch reionization
but also give us clues on the nature of these first sources.

As an example, we can consider the HE and the LE case. Although the
heating pattern and extent in both these cases are similar (see Figure
\ref{fig:tkin} \& \ref{fig:LEtkin}), the ionization profiles are
indeed different (see Figure \ref{fig:frach1} \&
\ref{fig:LEfrach1}). This according to equation \ref{eq:dtb} will be
reflected in the brightess temperature.

%
%
\begin{figure}
\centering
\hspace{-.4cm}
\includegraphics[width=.5\textwidth]{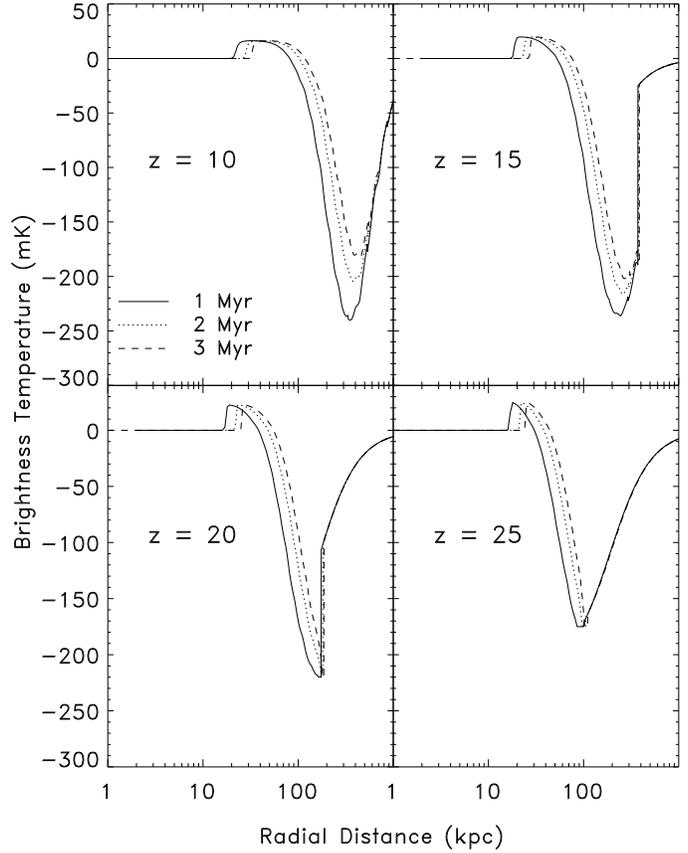}
\vspace{.3cm}
\caption{Radial profile of brightness temperature $dTb$ for the same black hole
  model as in fig.\ref{fig:LEtspin}.}
\label{fig:tbrig}
\end{figure}

Figure~\ref{fig:tbrig} shows the radial profile for the expected
brightness temperature for the same case as discussed in the context of the
spin temperature above. For all cases there is almost a sudden jump in
the brightness temperature. This is due to the sharp transition in the
neutral fraction~(Ref Fig~\ref{fig:frach1}).

\begin{figure}
\centering
\hspace{-.1cm}
\includegraphics[width=.5\textwidth]{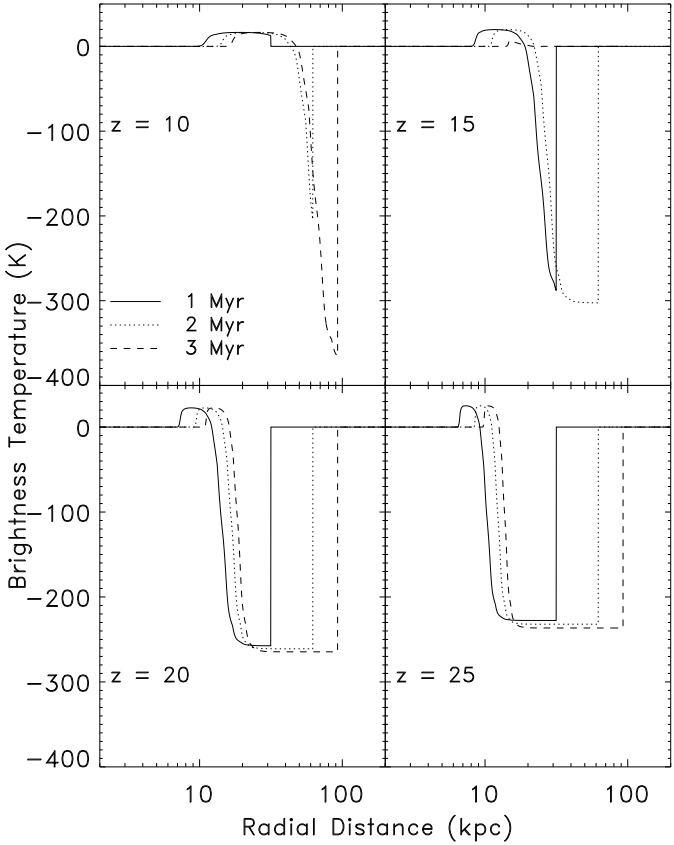}
\vspace{.5cm}
\caption{Brightness temperature calculated for the spin temperatures
of the star shown in fig.\ref{fig:startspin}. Assumption has been made
that the spin temperature remains coupled to the kinetic temperature
for a radius ``$ct$''.}
\label{fig:startbrig}
\end{figure}

The ionized fraction (Fig.\ref{fig:starfrach1}) and spin temperature
profile (Fig.\ref{fig:startspin}) is also reflected in the brightness
temperature in the case of stars as in Figure \ref{fig:startbrig}. The
maximum of the brightness temperature has a typical value between 20
and 30 mK for all redshifts but the negative wing spans different
ranges depending on the redshift.
%
%

An interesting observation is that the maximum value of $T_b$ remains the same
throughout the mass range. The reason being that in all cases $T_s$ is driven 
far away from $T_\rmn{CMB}$, such that the term $\left(1-\frac{T_\rmn{CMB}}
{T_s}\right)$ in equation \ref{eq:dtb} approximates to unity thus removing the
dependence of $T_b$ on $T_s$. As a consequence of this, we also see that the brightness temperature does not change considerably during the dorment period of the quasar. Figure \ref{fig:compareDENS} shows ionized fraction and temperature profile after the quasar has been switched off. But the ionized fraction is still low ($\approx 10^{-4}$) in the inner regions of the I-front, and the temperature is considerably high. Thus, the brightness temperature profile essentially remains the same. 

\citet{zaroubi05} have found that for power-law sources like
miniqsos the brightness temperature has an extended feature in the
transition from the ionized to neutral IGM. This feature has the same
order of magnitude of the one found here (see
Figure~\ref{fig:tbrig}). However, it should be noted that the origin
of this feature in both cases is completely different. While in
\citet{zaroubi05} the extended feature is produced by the
ionization-recombination balance of high energy photons, here the
origin of this feature is in the heating not ionization pattern around
these sources, also caused by high energy photons. As
Figure~\ref{fig:frach1} shows the neutral fraction of hydrogen changes
abruptly at the I-front location, namely, we here show that the
equilibrium assumption of \citet{zaroubi05} is never attained during
the miniqso lifetime.

 \subsection{Observability with future telescopes} As
mentioned before, there are several interferometric radio telescopes
underway to measure the 21cm signature of the EoR. It would be of
interest to investigate into the possiblity measuring directly the
topology of the 21cm through the brightness temperature. This of
course depends on the angular resolution, that measures the topology
on the sky and the frequency resolution that will provide the depth
information. We look into these aspects in the context of LOFAR and
SKA measuring the 21cm signal from redshift
10. LOFAR\footnote{www.astro.rug.nl/\~{ }LofarEoR} will have an
angular resolution of about 3-5 arcmin and a spectral resolution of
1MHz. This would correspond to a physical size on the sky of $\approx
1.4$ Mpc and $\approx 5$ Mpc in depth. On the other hand SKA with an
angular resolution of about 1 arcmin \citep{carilli04} would
correspond to $\approx .5$ Mpc on the sky. The spectral resolution for
the EoR experiment would more or less remain the same to beat other
noise effects. In the light of these rough estimates we realize that
it would be unlikely for LOFAR to see individual sources, although
once the ``bubbles'' start merging the sizes may become
comparable. But SKA, in principle should be able to resolve at least
the relatively big bubbles of the order of 100 of kpc easily.

\section{Sunyaev-Ze'ldovich effect}
\label{szsection}

Possibilities of observing the thermal and kinetic Sunyaev-Ze'ldovich
(SZ) effect during the epoch of reionization had been first
investigated by \citet{aghanim96} and more recently by
\citet{iliev07}.  A hot ball of ionized gas around first stars or
miniqsos resembles the classic case of clusters at lower redshifts in
which the SZ effect is observed today. Thus looking for signs of
reionization through the SZ effect is an obvious next step. A spin-off
of the simulations of the ionization and heating of the IGM by
miniqsos is the estimation of their contribution to the
Sunyaev-Ze'ldovich (SZ) effect. Both, the thermal and the kinetic
SZ-effects have been considered. 

For these simulations we start at z of 20 with two seed black hole masses,
$10^3$ and $10^4 M_\odot$. These two cases are
explored in order to reach, through accretion, masses of quasars
observed at redshift 6 by the SDSS. The black holes are allowed to grow
in time according the following equation.
\begin{equation}
M(t) = M_0\, \exp(f_{duty}{t/t_E}),
\label{bhmassgrowth}
\end{equation}
where $M(t)$ is the mass of the black hole at time '$t$', $f_{duty}$
the duty cycle of the black hole and $t_E$ the time scale of growth
which is around 41 $\times~(\epsilon_{rad}/0.1)$ Myrs. The radiation
efficiency of $\epsilon_{rad}$ was set to 0.05 \citep{loeb06}. We
assumed a duty cycle of 30\%.

First objects are formed in regions that are overdense with respect to
the IGM. Thus, we ran the above simulations for a density that is ten
times the mean IGM density at that epoch.


 \noindent\emph{Thermal SZ-effect}:

Hot electrons transfer energy to the CMB photons via inverse Compton 
scattering, redistributing the photons in the spectrum generally towards 
higher energy \citep{suny,sz}. The temperature fluctuation is thus given 
by;
\begin{equation}
\frac{\Delta T}{T}=-2\frac{k_b \sigma_T}{m_e c^2}\int_0^R T_e(t;l) n_e(t;l) dl
\label{sz}
\end{equation}
where $T_e$,$m_e$ and $n_e$ are the electron's temperature, rest mass and 
density, respectively. $\sigma_T$ is the Thomson cross section and $R$ is the 
radius of the ionized region. $T_e$ and $n_e$ are  functions of the radius and 
time after the quasar was switched on.  All parameters required to 
compute  equation~(\ref{sz}) have been obtained from the simulation. 
%
%
\begin{figure}
\centering
\hspace{-.6cm}
\includegraphics[width=.5\textwidth]{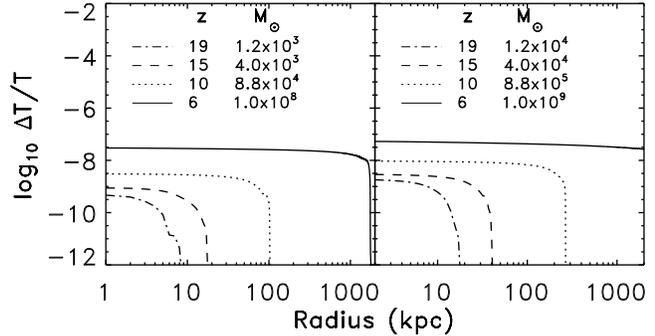}
\vspace{.4cm}
\caption{\emph{Thermal Sunyaev-Ze'ldovich effect:~} Radial profile of
temperature fluctuation $\frac{\Delta T}{T}$ for seed black hole
masses $10^3M_\odot$ and $10^4 M_\odot$ beginning at redshift $z=20$
are shown. The different lines corresponds to
particular redshifts as indicated in figure. Corresponding mass
according to equation \ref{bhmassgrowth} is shown.}
\label{fig:szthermal}
\end{figure}

The SZ-effect is the line of sight integral of the electron
pressure. In order to estimate the effect we considered the profiles
of the temperature and the ionized fraction to be spherically
symmetric and then projected the effect onto the plane of the
sky. Figure \ref{fig:szthermal} shows the temperature fluctuations
resulting from the black hole model mentioned above. Results are
plotted for different redshifts along with their corresponding masses,
down to a redshift of 6. All of them end at redshift 6 with masses
comparable to the black hole masses estimated for the SDSS high
redshift quasars~\citep{fan03, fan06}.  Massive miniqsos can cause
temperature fluctuations at high redshifts on the order of $10^{-7}$K.


\noindent\emph{Kinetic SZ-effect}
 
Apart from the thermal SZ-effect, there is
also the possibility of observing these large ionized bubbles through
the kinetic SZ-effect. Peculiar Motion of an ionized bubble with
respect to the background leads to an additional change of radiation
temperature in its direction, because of the finite optical depth
associated with the bubble \citep{suny}. The average anisotropy
associated with an ionized bubble is given by,
\begin{equation}
\frac{\Delta T}{T}=\frac{2}{3}\frac{v_r}{c}\int_0^R 2\sigma_T n_e(t;l) dl
\label{szkin}
\end{equation}
where $v_r$ is the radial component of the peculiar velocity of the ionized
bubble. The electron density $n_e(t;l)$ is given by the simulation and the 
radial component of the velocity dispersion is computed in the limit of linear
theory and is given by,
\begin{equation}
v_r(z) = v_r(0)(1+z)^{-1/2},
\label{veldisp}
\end{equation}
where $v_r(0)$ is velocity dispersion at redshift zero fixed at 600 km/s.
%
%
\begin{figure}
\centering
\hspace{-.6cm}
\includegraphics[width=.5\textwidth]{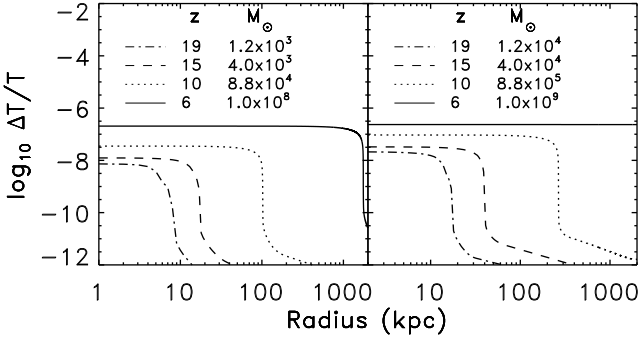}
\vspace{.4cm}
\caption{\emph{Kinetic Sunyaev-Ze'ldovich effect:~}
Radial profile of temperature fluctuation ${\Delta T}/{T}$ for the same model as in
figure \ref{fig:szthermal}}
\label{fig:szkin}
\end{figure}

The temperature fluctuation as a result of the kinetic SZ effect
(eq.~\ref{szkin}) is shown as a function of the radial distance from
the source in figure \ref{fig:szkin}. Again, as for the previous case,
we assumed spherical symmetry. The kinetic SZ- effect is in fact
larger in magnitude than thermal SZ-effect, making it easily visible
on the temperature maps of PLANCK.  Therefore, if the sizes and
temperatures of these early ionized bubbles are of the same order of
magnitude predicted here, then the SZ-effect is an important and
independent probe of the impact of early sources on the reionization
history that could be observed in the near future.

\section{Summary and Conclusions}
\label{conclusions}
In this paper a 1-D radiative transfer code has been developed in
order to study in detail the influences of primordial black holes and
Pop III stars on their surrounding environment.  We have shown, in
agreement with many other authors, that black holes and/or Pop III
stars have the potential to have been the primary source of
ionization. Although the ionized regions around typical black holes
and Pop III stars is not shown to be convincingly different modulo the
resolution of the telescopes like LOFAR, the heating around these two
sources do have a significantly different profile. This in turn is
reflected in the spin temperature of the system, which then directly
translates into the brightness temperature measured by the radio
telescope.

Spin temperatures are coupled to the kinetic temperature either through 
collisions or through Ly$\balpha$ pumping. Strengths of these coupling terms
are studied as a function of the radial distance from the source. 
Ly$\balpha$ photons do couple the spin temperature efficiently to the 
kinetic temperature hence making the brightness temperature insensitive to the 
background CMB, at least for a substantial distance ($\approx 2-3 Mpc$) away 
from the source. 

Recombination timescales are orders of magnitude lower than that of
ionization.  Therefore, an HII bubble remains ionized for a long
period after the source of radiation has been switched off. Also the
temperature remains largely unchanged for a significant fraction of
Hubble time since cooling processes typically involve two body
interaction that increase the timescales involved. Interestingly for
miniqsos, a marginal increase in the ionized bubble after the
switching off of the central ionizing source is observed. Collisional
ionizations are the dominant cause of this phenomenon which is
therefore relatively less in stars because of the lack of high energy
X-ray photons which heats up the environment significantly hence
boosting the collisional ionizations.

For miniqsos with a hard spectrum that lacks the UV part of the
spectrum we observe a excess of HI fraction as a function of radius
just before the main ionization front. This is interpreted as a result
of the interplay between the increase of HeII and decrease of HeIII,
which increase the HI recombination rate,on the one hand and the
decrease in the photon flux as a function of radius on the other. This
interpretation is supported by the weakening of the HI excess feature
as a function of black hole mass. This phenomenon is not observed in
miniqsos with spectrum that has UV photons, which is due to the
ionization efficiency of these photons.

The results of the IGM heating around miniquasars have also been
compared to the analytical approach of \citet{zaroubi07}. Results
shown in Figure \ref{fig:compare} demonstrate that the analytical
solutions, while in agreement to within an order of magnitude, underestimate
the kinetic temperature away from the centre by roughly a factor of
up to 5. The difference in the results is due to the helium
species cross-sections are not being fully accounted for in
the analytical approach. In fact, in order for the heating to be accurately
calculated one needs an exact knowledge of the abundance of each of
the hydrogen and helium species averaged over their evolution history
until the point in time in which one is interested and such detailed
knowledge could not be obtained with the analytical approach.

Also, the thermal and kinetic Sunyaev-Ze'ldovich effects around
these quasars were studied. Two cases are considered in which mass
growth of the black hole is incorporated such that it winds up with
$10^{8}$ and $10^9 M_\odot$ black holes at redshifts close to six. The
estimated values for the temperature fluctuations are within the
sensitivities of the future mission like PLANCK.

We conclude that the brightness temperature which in turn reflects the
underlying ionization and heating around a source is sensitive to many
factors. The spectral energy distribution of the source, range of
energies spanned by the photons of these sources, their clustering
properties, the photon escape fraction, the redshifts at which the
sources turn-on, their lifetimes, and many other complex feedback
mechanisms, to name a few that will influence the brightness
temperature, which is the observable we are after.  Thus, in order to
span this large parameter space of possibilities a 1D radiative
transfer code is very useful.

An important next step for the radiative transfer code described here
is to include line emissions and absorptions, especially that of the
Ly$\alpha$ line, in the radiative transfer equations. We will also
incorporate the rate equation for molecular hydrogen and its various
states. 

Finally, in the near future we plan to incorporate the results of this
study into the output of N-body simulations.  Such an approach will
allow us to produce quick 21 cm maps of the EoR, which then will be
used in conjunction with the simulations of galactic and extra-galactic
foregrounds, ionospheric models and LOFAR specific instrument
responses to generate ``dirty-maps'' of the EoR.

\section*{acknowledgements}
The authors thank B. Ciardi, A. Nusser , E. Ripamonti and M. Spaans
for discussion and helpful comments. We are also thankful to
the anonymous referee for his illustrative and constructive
comments.

\appendix
\section{Analytical Approximation}
\label{analytical}
A number of papers have outlined analytical approximations for the
ionized fraction and heating of the IGM by miniquasars (cf.,
\citet{madau,zaroubi07}). In order to test the validity of these
analytical solutions, we compare such a solution with the RT
simulations for 100 and $10^4 M_\odot$ black hole masses at z of 20
and 10 and with a lifetime of 3 and 10 Myr. Here we follow the
analytical equations of \citet{zaroubi07} who calculate the ionization
species by solving the following ionization-recombination
equilibrium equation;
\begin{equation}
\alpha_{H I}^{(2)} n_H^2 (1-x_{H I})^2= \Gamma(E;r)~n_H x_{H I}
\left(1+{\sigma_{He}\over\sigma_{H} }{n_{He}\over n_{H}}\right).
\label{eq:balance}
\end{equation}
where $\Gamma(E;r)$ is the ionization rate per hydrogen atom for a
given photon energy at distance $r$ from the source. $\Gamma$ is
calculated as a function of $r$;
\begin{equation}
\Gamma (E;r)  =  \int_{E_0}^{\infty}\sigma(E) {\cal N}(E;r)\frac{dE}{E}.
\label{eq:gamma}
\end{equation}
$\alpha_{H I}^{(2)}$ is the recombination cross-section to the second
excited atomic level and has the value $2.6 \times
10^{-13}T_4^{-0.85} \mathrm{cm^3 s^{-1}}$, with $T_4,$ the gas
temperature in units of $10^{4~} K,$. For this calculation they assume
 $T=10^{4~}\mathrm{K}$. Although not very accurate, it gives a
lower limit on the recombination cross-section, $\alpha_{H I}^{(2)}$
(in neutral regions atomic cooling prevents the gas from having a
higher temperature).

\begin{figure}
\setlength{\unitlength}{1cm} \centering
\begin{picture}(8,9.4)
\put(-1.,-0.5){\includegraphics{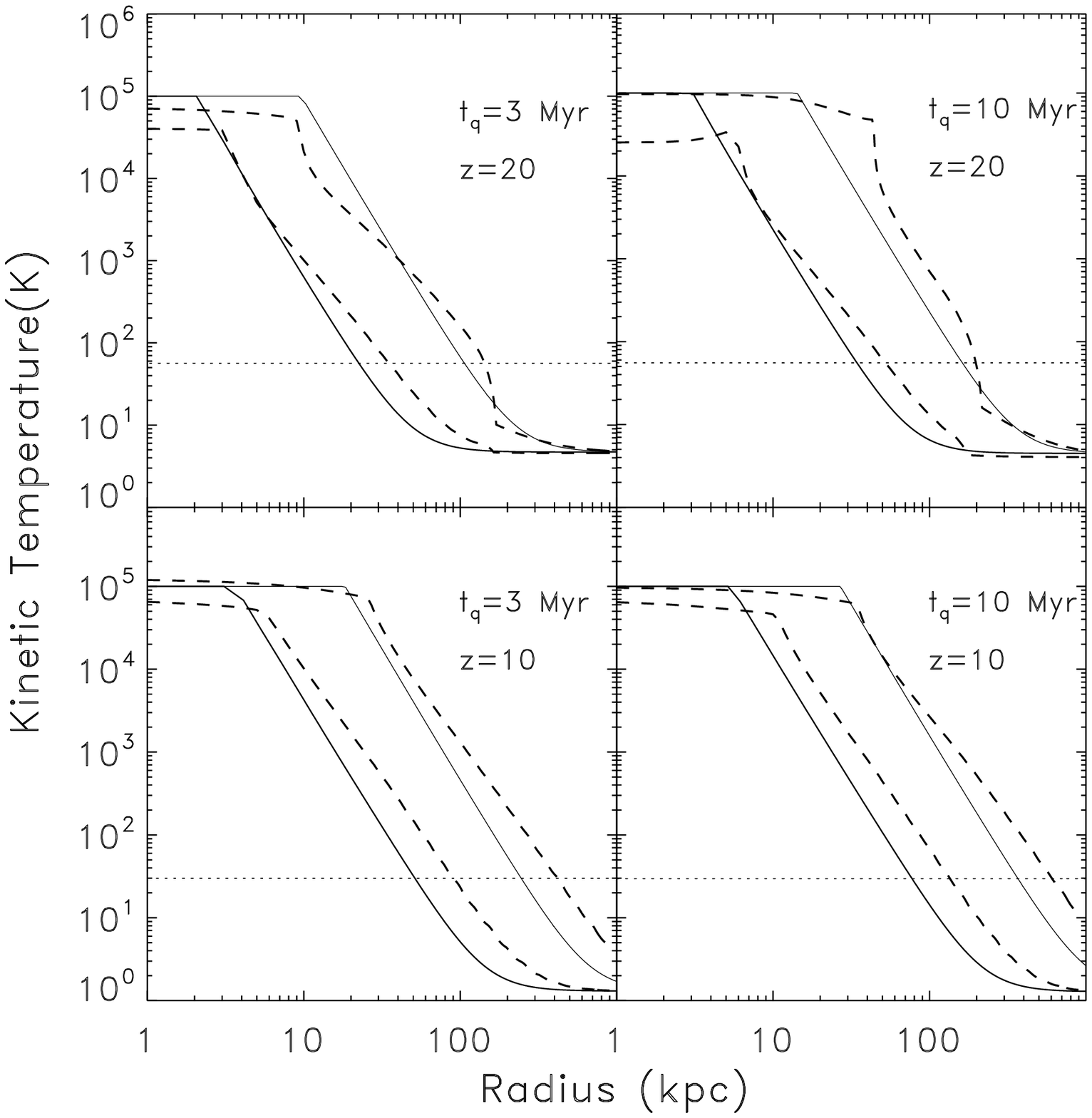}}
\end{picture}
\caption{Each panel in this Figure shows a comparison between the
model of \citet{zaroubi07} and the results of the spherically
symmetric radiative transfer code applied to black hole masses, 100 \&
10000 $M_\odot$ with the same spectral energy distribution. The
analytical calculation is represented by the solid line and that
obtained from the radiative transfer code is represented by the dashed
line. The different panels corresponds to different values of redshift
and quasar lifetime '$t_q$'}
\label{fig:compare}
\end{figure}

Similarly the heating rate ${\cal H}(r)$, is calculated as a function
of the radial distance $r$ using;
\begin{equation}
\label{eq:heat}
{\cal H}(r)  =  f n_{H} x_{HI}\int_{E_0}^{\infty}\sigma(E) {\cal N}(E;r) dE
\end{equation}
where $f$ is the fraction of the absorbed photon energy that goes into
heating through collisional excitations of the surrounding material
\citep{shull}. The function $f$ is fitted in the paper by \citet{shull} 
with the following simple fitting formula: $f = C \left[ 1- \left( 1-
x^a\right)^b \right]$, where $C=0.9771$, $a=0.2663$ and $b=1.3163$ and
$x=1-x_\rmn{HI}$ is the ionized fraction. This fitting function is valid
in the limit of high photon energies, an appropriate assumption for
the case at hand. The fitting formula is modified by imposing a lower
limit of $11\%$ for the fraction of energy that goes into heating as
the proposed fitting formula does not work well at ionized hydrogen
fractions smaller that $10^{-4}$. And $\sigma(E)$ is the cross-section
combining that of hydrogen and helium in the following manner;
\begin{equation}
\sigma(E) = \sigma_{H}(E)+\frac{n_{He}}{n_H}\sigma_{He}(E) = \sigma_1{\left(\frac{E_o}{E}\right)}^3.
\label{eq:hhecross}
\end{equation}
Here, $\sigma_1$ is a smooth function of energy. The temperature of the 
IGM due to this heating is determined by the following
equation \citep{madau, zaroubi07}:
\begin{equation}
 \frac{3}{2}\frac{n_H k_b T_{kin}(r)}{\mu} = {\cal H}(r) \times t_q.
\end{equation}
Here $T_{kin}$ is the gas temperature due to heating by collisional
processes, $k_b$ is the Boltzmann constant, $\mu$ is the mean
molecular weight and $t_q$ is the miniquasar lifetime. This equation
assumes that the heating rate due to the absorption of X-ray photons
during the miniquasar lifetime is constant.  Given the miniquasars
lifetime of 3 and 10 Myr and Hubble time at the redshifts we are
interested in, cooling due to the expansion of the Universe can be
safely neglected. A $10^{5}\mathrm{K}$ cutoff on the gas kinetic
temperature due to atomic cooling is imposed. Figure~\ref{fig:compare} 
shows the kinetic temperature profiles for the 4 cases namely, redshifts 
of 20 and 10 and lifetimes of 3 and 10 Myrs.
%
%

\begin{figure}
\setlength{\unitlength}{1cm} \centering
\begin{picture}(8,9.4)
\put(-1.,-0.5){\includegraphics{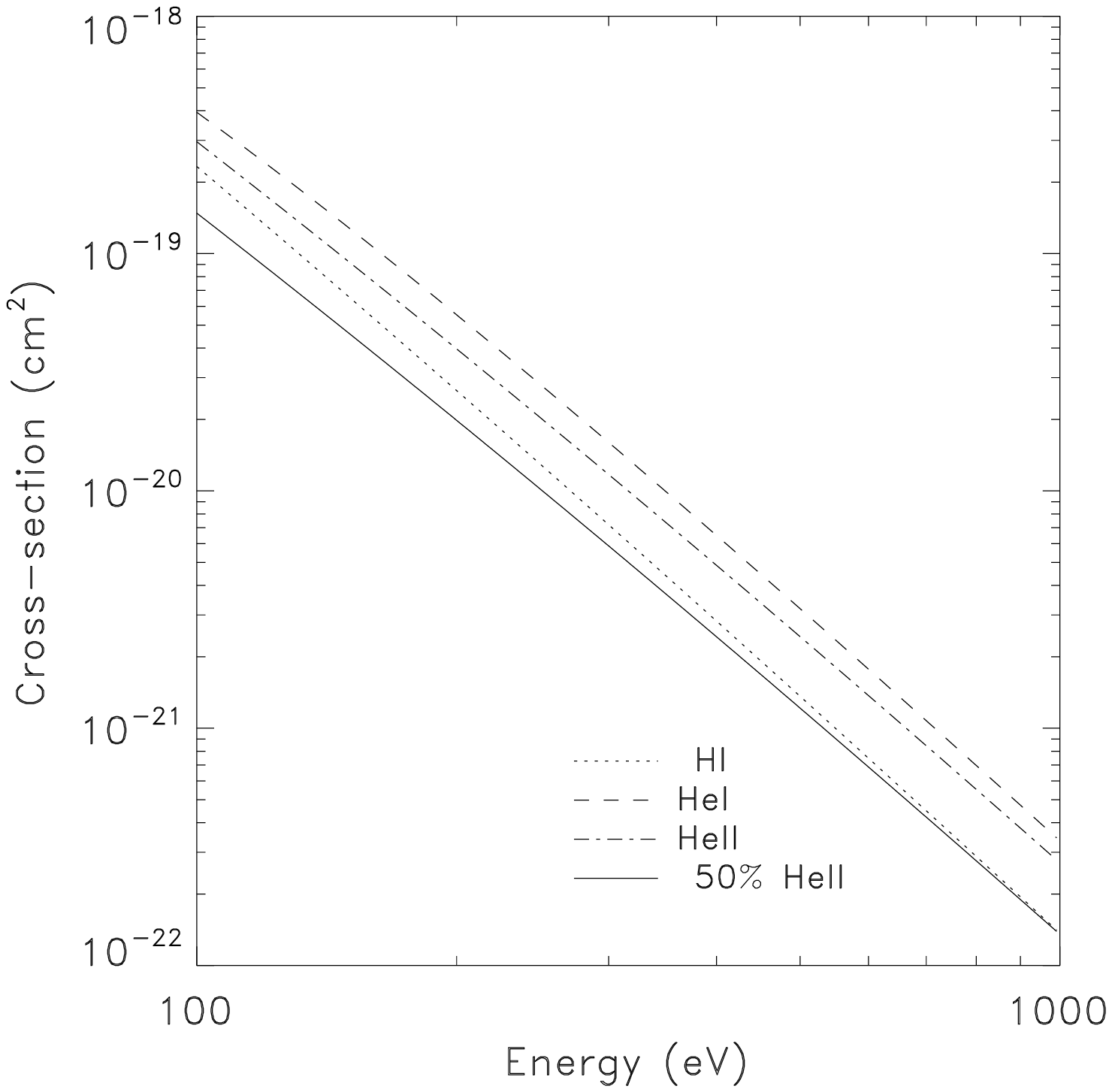}}
\end{picture}
\caption{ Figure shows the cross-sections \citep{verner} of neutral
hydrogen, neutral helium and singly ionized helium, all waited by
their abundances in the mean IGM. Over plotted is the also the 50\%
singly ionized helium, i.e., all hydrogen has been ionized and 50\% of
all helium is completely ionized.}
\label{fig:cross}
\end{figure}

Although the figure shows an overall agreement between the simplistic
analytical model and the full RT numerical solution, the analytical
approach underestimates the heating by a factor of roughly 3 to 5 in
the outer radii. The reason for this is that the helium cross-section
is accounted for in the analytical approach (see
eq. \ref{eq:hhecross}) in a very incomplete fashion whereby the
heating stops as soon as hydrogen is completely ionized (eq.
\ref{eq:heat}). In reality this is not the case as heating by high
energy photons will continue until also He is fully ionized. Figure
\ref{fig:cross} shows the cross-sections of neutral hydrogen (dotted
line), completely neutral helium (dashed line), singly ionized helium
(dot dashed line) all weighted by their abundances. Over plotted on
this is the case in which all hydrogen and half of all helium has been
completely ionized (50\% HeII case). At high energies, like the
spectrum of the miniqso under discussion, even this fraction of singly
ionized helium provides a substantial cross-section that can capture
the photon and convert its energy to heat. Interestingly, looking at
Figures \ref{fig:fracmix} and \ref{fig:LEfracmix} in conjunction with
Figures \ref{fig:tkin} and \ref{fig:LEtkin} around a radial distance
of 20 - 30 kpc reveals that there is a correlation between the increase
in temperature and the corresponding increase in the singly ionized
helium species.

%
%

\label{lastpage}

\end{document}